\documentclass[twocolumn]{aastex631}

\usepackage{longtable}
\usepackage{tabu}
\usepackage{lineno}

\begin{document}

\title{{Evaluating Short-Warning Mitigation via Intentional Robust Disruption of a Hypothetical Impact of Asteroid 2023 NT1}}

\author[0009-0004-7968-3859]{Brin Bailey}
\affiliation{University of California, Santa Barbara, Department of Physics, Santa Barbara, CA 93106, USA}
\email{brittanybailey@ucsb.edu}

\author[0000-0003-2796-251X]{Alexander N. Cohen}
\affiliation{University of California, Santa Barbara, Department of Physics, Santa Barbara, CA 93106, USA}

\author[0000-0002-6443-7303]{Sasha Egan}
\affiliation{New Mexico Institute of Mining and Technology, Energetic Materials Research and Testing Center, Socorro, NM 87801, USA}

\author{Philip Lubin}
\affiliation{University of California, Santa Barbara, Department of Physics, Santa Barbara, CA 93106, USA}

\author{Ruitao Xu}
\affiliation{University of California, Santa Barbara, Department of Physics, Santa Barbara, CA 93106, USA}

\author{Mark Boslough}
\affiliation{University of New Mexico, 1700 Lomas Blvd, NE. Suite 2200, Albuquerque, NM 87131, USA}

\author{Darrel K. Robertson}
\affiliation{NASA Ames Research Center, Moffett Field, CA 94035, USA}

\author{Elizabeth A. Silber}
\affiliation{Sandia National Laboratories, Albuquerque, NM 87123, USA}

\author{Irina Sagert}
\affiliation{Los Alamos National Laboratory, Los Alamos, NM 87544, USA}

\author{Oleg Korobkin}
\affiliation{Los Alamos National Laboratory, Los Alamos, NM 87544, USA}

\author{Glenn Sjoden}
\affiliation{University of Utah, Salt Lake City, UT 84112, USA}



\begin{abstract}

We investigate various short-warning mitigation scenarios via fragmentation for a hypothetical impact of asteroid 2023 NT1, a Near-Earth Object (NEO) that was discovered on July 15, 2023, two days after its closest approach to Earth on July 13. The asteroid passed by Earth within $\sim$0.25 lunar distances, with a closest approach of $\sim$$1\times10^5$ km and velocity of 11.27 km/s. Its size remains largely uncertain, with an estimated diameter range of 26 – 58 m and a most probable estimate of 34 m\footnote{JPL Sentry, September 15, 2023} (weighted by the NEO size frequency distribution). If 2023 NT1 had collided with Earth, it could have caused significant local damage. Assuming a spherical asteroid with a diameter of 34 m, uniform density of 2.6 g/cm$^3$, and impact velocity of 15.59 km/s, a collision would have yielded an estimated impact energy of $\sim$1.5 Mt, approximately three times the energy of the Chelyabinsk airburst in 2013. We analyze the effectiveness of mitigation via intentional robust disruption (IRD) for objects similar to 2023 NT1. We utilize Pulverize It (PI), a NASA Innovative Advanced Concepts (NIAC) study of planetary defense via fragmentation, to model potential mitigation scenarios through simulations of hypervelocity asteroid disruption and atmospheric ground effects in the case of a terminal defense mode. Simulations suggest that PI is an effective multi-modal approach for planetary defense that can operate in extremely short interdiction modes, in addition to long interdiction time scales with extended warning. Our simulations support the proposition that threats like 2023 NT1 can be effectively mitigated with intercepts of one day (or less) prior to impact, yielding minimal to no ground damage.

\end{abstract}


\section{Introduction} \label{sec:intro}
\subsection{Initial characterization of 2023 NT1} \label{sec:2023 NT1 params}

Little is known with certainty about asteroid 2023 NT1, and it is important to note that most parameters described here are estimates based upon very few well-constrained measurements. 2023 NT1 is classified as an Apollo Near-Earth Asteroid (NEA) and was first observed on July 15, 2023 using the M22 ATLAS-Sutherland telescope in South Africa \citep{noauthor_iau_nodate}. The asteroid made its closest approach to Earth on July 13, 2023 at 10:13 ($\pm$ 1 minute) TDB and passed at a distance of $6.7\times10^{-4}$ au ($\sim$10$^5$ km) (determined via a 3$\sigma$ body target-plane error ellipse) with a velocity of 11.27 km/s \citep{noauthor_small-body_nodate}. If 2023 NT1 had impacted Earth on July 13, rather than narrowly missing, it is estimated that it would have had an impact velocity (the velocity at atmospheric entry) of 15.59 km/s \citep{noauthor_sentry_nodate}. This impact velocity was approximated from the asteroid's observed close approach velocity with additional contribution from Earth's gravitational potential energy.  The body’s absolute magnitude ($H$; defined as the apparent magnitude at 1 au from the Sun and observer) was observed as 25.05 ($\sigma$ = 0.37706) \citep{noauthor_small-body_nodate}. 

While the body’s closest approach distance and velocity and absolute magnitude are relatively constrained, the remainder of orbital and physical characteristics of 2023 NT1 are not known with certainty. Observations of 2023 NT1 made by NASA's Jet Propulsion Laboratory (JPL) and the International Astronomical Union (IAU) from July 15-September 16, 2023 yielded an eccentricity of 0.51271 $\pm$ 1.4427$\times$10$^{-4}$  with an estimated orbital period of 949.58 $\pm$ 0.03919 days (2.5998 $\pm$ 1.0729$\times$10$^{-4}$ years) \citep{noauthor_small-body_nodate}. At its closest approach, the asteroid was about 35 days past perihelion on its orbit around the Sun with a semi-major axis of 1.8907 \textcolor{blue}{$\pm$ $5.2018\times10^{-4}$} au and an inclination of 5.7712 $\pm$ 5.2018$\times$10$^{-4}$ degrees \citep{noauthor_small-body_nodate}. 

Diameter estimates of the object are based on its absolute magnitude, yielding a size range of 26-58 m \citep{noauthor_small-body_nodate,noauthor_sentry_nodate}. By assuming a uniform spherical body with visual albedo p$_v$=0.154 (in accordance with the Palermo Scale) and weighting by impact probability and the NEO size frequency distribution, JPL Sentry estimates that 2024 YR4 has a probable diameter of 55 m, thought to be accurate to within a factor of 2 \citep{noauthor_sentry_nodate}. Assuming a spherical body with a 34 m diameter and uniform density of 2.6 g/cm$^3$, the mass is roughly 5.2$\times$10$^7$ kg, which is thought to be accurate to within a factor of 3 \citep{noauthor_sentry_nodate}. Assuming these characteristics, 2023 NT1’s mean impact energy release is equivalent to $\sim$1.5 Mt of TNT. 

\subsection{Earth impact risk} \label{sec:impact risk}

On September 16, 2023, JPL Sentry removed 2023 NT1 from its Impact Risk page, as additional observations ruled out the possibility of a future impact  \citep{noauthor_sentry_nodate}. However, the conditions that allowed 2023 NT1’s close passage without detection indicate that existing planetary defense programs are inadequate for addressing short-term threats. 

In relation to historical events, the diameter estimates of 2023 NT1 place the body between the size range of the Chelyabinsk asteroid of 2013 (18 m diameter) and the Tunguska impactor of 1908 (estimated 40-50 m diameter). Both events caused disruptions to local human life and land. The Chelyabinsk airburst released energy equivalent to 0.57 $\pm$ 0.15 Mt of TNT, injuring over 1,600 people and damaging more than 7,000 buildings, resulting in an estimated \$33 million (1 billion rubles) in damages \citep{popova_chelyabinsk_2013,brown_flux_2002,noauthor_five_nodate,noauthor_russian_nodate}. The Tunguska event produced an estimated TNT equivalent yield of 3-50 Mt (most likely 3-15 Mt), which flattened a large portion of the surrounding forest (potentially on the scale of millions of trees) and resulted in sparse fires extending up to 10-15 km from the epicenter \citep{popova_chelyabinsk_2013,brown_flux_2002,jenniskens_tunguska_2019,longo_tunguska_2007}. Objects at least the size of the Chelyabinsk asteroid are expected to impact Earth approximately every 50-100 years, while objects with the energy of the Tunguska asteroid are expected to impact Earth approximately every 300-1,000 years ($\sim$300-500 years for an estimated 3-5 Mt event; $\sim$800-1,800 years for an estimated 10-15 Mt event) \citep{harris_population_2015,boslough_low-altitude_2008,brown_flux_2002}. 

\subsection{Threat detection} \label{sec:detection}

While 2023 NT1 currently poses no threat to Earth, it is conceivable that similar objects could go undetected, leading to short-warning or no-warning impacts. A critical component of any mitigation system is the ability to detect threats promptly. While detection rates for large threats (\textgreater1 km) are reasonably fulfilled, our situational awareness for the smaller asteroids (\textless140 m diameter), which constitute the most common threats in the solar system, is poor. As of September 2024, it is estimated that 94\% of NEAs \textgreater1 km in diameter have been discovered, for which there is an estimated population of about 940 \citep{noauthor_near-earth_nodate,harris_population_2021}. In comparison, estimates suggest that approximately 44\% of NEAs \textgreater140 m have been discovered, for which there is an estimated population of about 25,000 \citep{noauthor_near-earth_nodate}. For NEAs \textless140 m, a 2021 study by Harris and Chodas estimate a population of approximately 2.88$\times$10$^{9}$, though this value has a large degree of uncertainty \citep{harris_population_2021}. As asteroid diameter decreases, the likely population increases drastically, thus decreasing our estimated discovery ratio.

Given the power law of threat incidence versus diameter and a lower threshold of approximately 20 m in diameter for what are considered ``significant" threats (for rocky compositions), our situational awareness is severely limited in the smaller threat regime. This issue may have two solutions, distinct in approach but aligned in goal: 1) enhancing small-threat situational awareness through heightened detection efforts, and 2) augmenting it by establishing effective short-term mitigation systems. Both strategies could theoretically be enhanced concurrently to accelerate the development of a robust planetary defense program. This topic is discussed further in Section \ref{sec:PD limitations}.


\section{PI for planetary defense} \label{sec:pi}
\subsection{Introduction to PI} \label{sec:pi method}
Pulverize It (PI) is a NASA Innovative Advanced Concepts (NIAC) Phase II study of planetary defense which is intended to operate in both short-warning and extended interdiction modes, representing a fundamentally different approach to threat mitigation. Planetary defense has traditionally focused on mitigation via orbital modification, or deflection, utilizing momentum transfer to prevent an impact. Mitigation via deflection has explored a wide  range of possible techniques ranging from impulsive methods, like direct impact or nuclear ablation \citep{rivkin_double_2021,dearborn_options_2020}, to gradual orbit deflection (e.g., via surface albedo alteration) \citep{hyland_permanently-acting_2010}, to the utilization of gravity tractors, ion engines, laser ablation, and other developing technologies \citep{mazanek_enhanced_2015,walker_concepts_2005}. Alternatively, PI uses energy transfer for mitigation, aiming to fragment a threat via intentional robust disruption (IRD) rather than deflect it. The method utilizes an array of hypervelocity kinetic penetrators that disassemble an asteroid into many small (typically \textless10 m) fragments \citep{lubin_asteroid_2023}. 

Depending on the time scale of interception, the fragment cloud either misses Earth entirely (long-warning time) or is dissipated in Earth’s atmosphere (short-warning time, henceforth referred to as the ``terminal mode”) \citep{lubin_asteroid_2023}. The latter results in a series of airburst events with spatial and temporal spread at varying high altitudes, which distributes the energy of the parent asteroid \citep{lubin_asteroid_2023, bailey_optical_2024}.

Note that PI's preferred mode of operation is its extended interdiction (long-warning) mode, where a threat is intercepted months or years ahead of Earth impact (depending on threat size and velocity, among other parameters). We intend for the use of PI in its terminal mode to be reserved for extreme cases where there is little to no time before impact, in which there would be no other mitigation option. When possible, it is preferable to keep any planetary defense operation as far from Earth as possible to mitigate any potential damage on the ground. As such, note that the methodology and simulated scenarios presented here focus only on ``worst-case" scenarios. A more thorough discussion of PI in its extended warning and enhanced deflection modes can be found in Lubin and Cohen \citep{lubin_asteroid_2023}.

\subsection{Terminal mitigation} \label{sec:mitigation}
Mitigation of a threat by using PI in a terminal mode consists of two stages: 1) interception and fragmentation of the asteroid and 2) dissipation of the fragment cloud within Earth’s atmosphere. In such scenarios, the primary mechanism for threat mitigation is the distribution of the asteroid’s energy into the fragment cloud, resulting in spatially and temporally de-correlated ground effects. 

In the terminal mode, the impacting fragment cloud interacts with Earth's atmosphere in a manner similar to an unmitigated asteroid airburst, but instead disperses the energy relative to the unmitigated case \citep{lubin_asteroid_2023}. During atmospheric entry of the fragments, the high-velocity ram pressure (or stagnation pressure) exerted by the atmosphere eventually exceeds the material yield strength, initiating a cascading breakup event \citep{kring_chelyabinsk_2014,robertson_effect_2017}. The ram pressure is determined by the density of the atmosphere and velocity of the parent asteroid, whereas yield strength depends largely on the shear strength provided by the internal structure and integrity of the asteroid, including the strength of its components (as well as other parameters such as size, density, velocity, and entry angle) \citep{robertson_effect_2017}. As the pressure buildup on the fragment increases, it undergoes ablation and begins to flatten and expand, increasing the surface area on which the rising aerodynamic drag can act \citep{kring_chelyabinsk_2014}. This runaway process eventually converts the fragment’s kinetic energy into a release of heat and pressure through detonation, or ``bursting," of the fragment \citep{kring_chelyabinsk_2014}. 

Depending on the material strength, initial failure can occur either externally or internally; the failure site will influence the method by which the fragment bursts \citep{robertson_effect_2017}. These airbursts yield optical pulses and de-correlated shock waves on the ground (hereafter referred to as ``ground effects") which, in reasonable mitigation scenarios that are appropriate for the threat, result in little to no damage.

\subsection{Optical and acoustic ground effects}\label{sec:ground effects thresholds}
The ground effects of unmitigated asteroids or large fragments ($>$20 m) have the potential for significant destruction; we therefore design mitigation scenarios such that the fragments are generally $<$10 m in diameter \citep{lubin_asteroid_2023}. It is important to analyze both the optical and acoustic ground effects to design proper mitigation scenarios with acceptably low damage.

For optical damage, there are two key effects to highlight: total energy deposition (J/$\text{m}^2$) and time-dependent power output (W/$\text{m}^2$). High energy deposition can lead to effects such as fires and skin damage (sunburn), whereas high power outputs can cause temporary to permanent blindness \citep{glasstone_effects_1977}. We set an optical energy damage threshold of 200 kJ/$\text{m}^2$ so that the sum of the optical energy taken over all fragments (assuming the analytic relationship outlined in Section \ref{sec:optical ground effects}) is kept below this value. We chose a value of 200 kJ/$\text{m}^2$ ($\sim$5 cal/cm$^2$ of radiant exposure), as this is the point at which combustible organic materials (like leaves and paper) can begin to catch fire \citep{martin_diffusion-controlled_1965}.

For acoustic effects, studies of window damage in atmospheric nuclear tests \citep{glasstone_effects_1977} have found that the threshold for residential window breakage corresponds to peak pressures of about 3 kPa. PI therefore sets a shock wave over-pressure threshold of 3 kPa, with the goal in any mitigation scenario being to keep all shock wave over-pressures (including constructive interference) below this value.

\begin{figure*}
    \centering
    \includegraphics[width=0.9\textwidth]{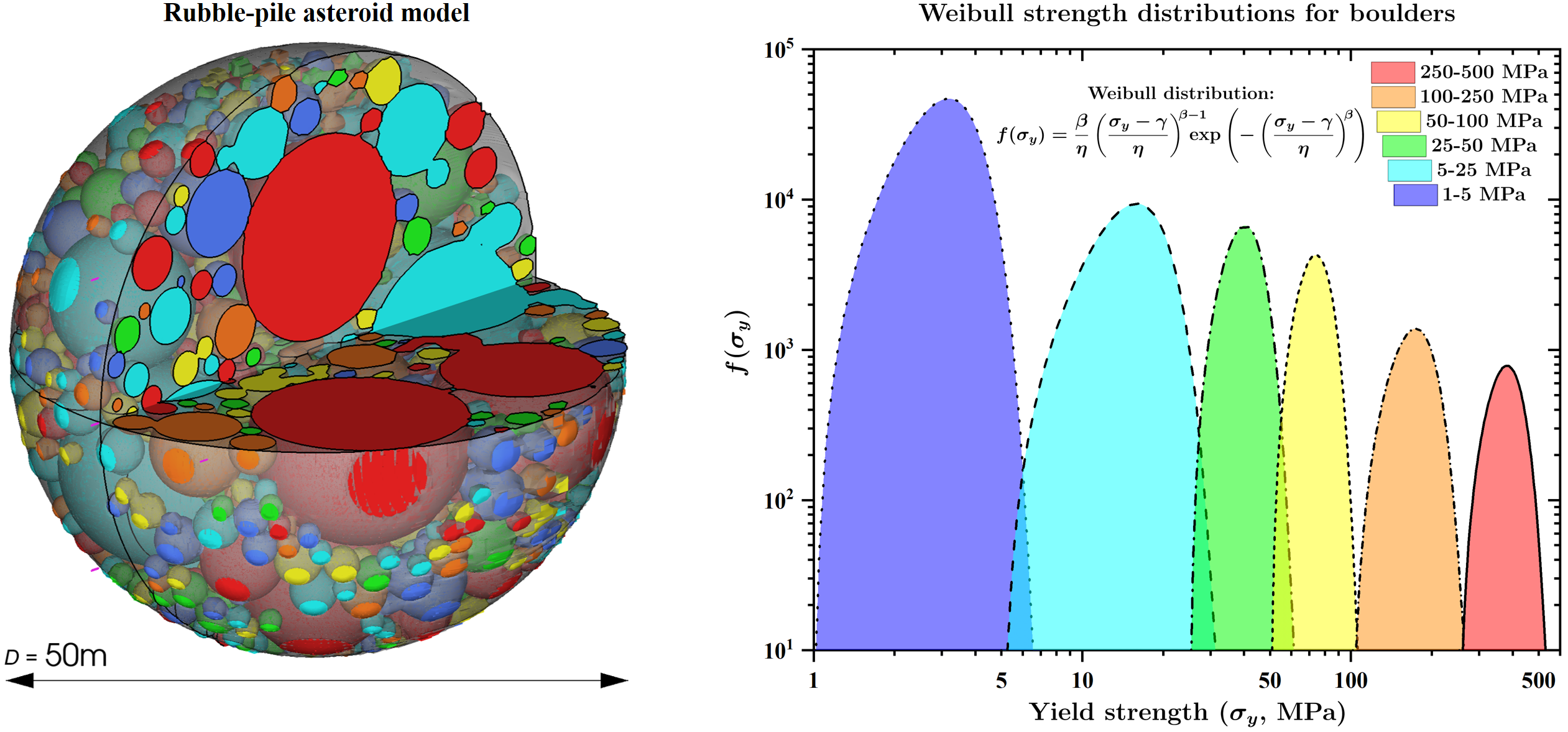}
    \caption{(Left) Example rubble pile asteroid model shown in partial cross section for a 50 m diameter bolide. The binder material is shown in transparent grey and the boulder distribution within it is colored by the material yield strength ranges as shown on the right in the plot of the Weibull strength distributions. (Right) Weibull strength distributions for the six boulder types in our rubble pile asteroid models. These distributions are normalized and used to initialize the yield strengths of the boulders, scaling up from 1 MPa initial yield strength. Additionally, a seventh distribution is assigned to the binder material, with a weak strength distribution ranging from approximately 1--50 Pa. For reference, the violet 1–5 MPa distribution is comparable to hardened soil, the cyan 5–25 MPa distribution to standard grade concrete, the green 25–50 MPa distribution to high strength concrete, the yellow 50–100 MPa distribution to aluminum, the orange 100–250 MPa distribution to structural steel, and the red 250–500 MPa distribution to high strength steel and titanium. These strengths are an extremely conservative overestimation of the strength of rubble pile asteroids \citep{sanchez_strength_2014,pravec_fast_2000}. This figure from \citep{cohen_asteroid_2024} is licensed under \href{https://creativecommons.org/licenses/by-nc-nd/4.0/deed.en}{CC BY-NC-ND 4.0}.}
    \label{fig:weibull}
\end{figure*}

\begin{figure*}
    \centering
    \includegraphics[width=0.9\textwidth]{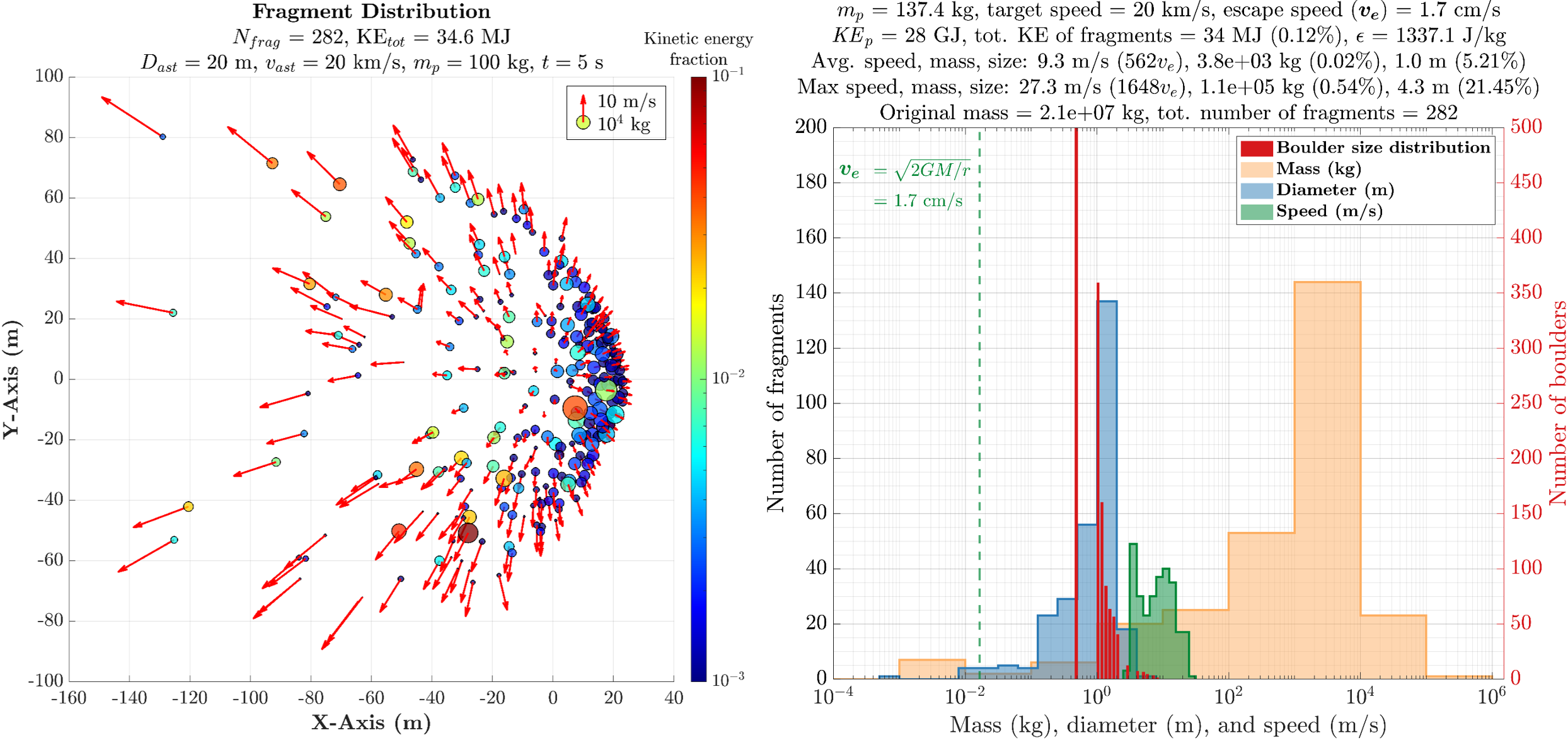}
    \caption{Fragment distribution and statistics at $t$ = 5 seconds after impact for a 100 kg 10:1 aspect ratio tungsten cylindrical penetrator incident at 20 km/s upon a 20 m diameter rubble pile asteroid target, as discussed in Section \ref{sec:20m intercept}. The orange, blue, and green histograms indicate the distributions of fragment masses in kilograms, diameters in meters, and velocity in meters per second, respectively. Of note is the average fragment size of 1 m and average fragment velocity of 9.3 m/s, which is more than 1000 times the the gravitational escape velocity of 1.7 cm/s (dashed green line on left). Also note that the maximum fragment size is 4.3 m, which is well below the 10 m acceptable fragment size threshold for rocky asteroid densities around 2.6 g/cm$^3$ \citep{lubin_asteroid_2023}. These results suggest that asteroid 2023 NT1, if similar to the Chelyabinsk asteroid in size and composition, could be mitigated using a single tungsten penetrator with mass on the order of 100 kg, assuming an asteroid closing velocity of 20 km/s. This figure from \citep{cohen_asteroid_2024} is licensed under \href{https://creativecommons.org/licenses/by-nc-nd/4.0/deed.en}{CC BY-NC-ND 4.0}.}
    \label{fig:20m impact sim}
\end{figure*}

\section{Interception and fragmentation} \label{sec:intercept}

The dynamics of a hypervelocity intercept with the hypothetical threat were simulated using the Lawrence Livermore National Laboratory’s (LLNL) Arbitrary Lagrangian–Eulerian three-dimensional (ALE3D) hydrodynamics code \citep{noauthor_ale3d_nodate}. The impact was modeled in 3D using Livermore Equation-of-State (LEOS) material models, which include material shock response, vaporization, and ionization effects. The interception process appears to rapidly convert the penetrator’s kinetic energy (within the asteroid reference frame) into heat and shock waves. The energy of the penetrator impact is sufficient to locally vaporize and ionize material near the impact site, while the generated shock waves damage and fracture the asteroid material as they propagate, reflect, and refract. The expanding region of vaporized material imparts enough energy to the bulk of the asteroid to drive the fragments apart with sufficient kinetic energy to overcome the asteroid’s gravitational binding energy. 

\subsection{Penetrator model}

One of the primary challenges in designing an ALE simulation is balancing the desired level of accuracy with computational feasibility. To make these calculations manageable, the region of interest is divided into smaller, discrete units called ``cells.” This division forms a ``mesh,” which serves as the framework for numerical computations. The size of the mesh and the number of elements within it directly impact both the precision and efficiency of the simulation. Finer mesh resolutions provide greater precision, but demand significantly more computational power and time. Conversely, coarser mesh resolutions reduce computational demands at the expense of accuracy.

A useful concept in managing this trade-off is the distinction between ``idealized” and ``realized” masses. \textit{Idealized masses} represent the theoretical values based on the model’s input parameters. In contrast, \textit{realized masses} are the actual values computed within the simulation, which can be influenced by the mesh resolution and computational constraints. Understanding and managing the relationship between these two values is critical for achieving reliable simulation results.

For this study, we used a tungsten penetrator modeled as a 10:1 aspect ratio cylinder with a density of 19.24 g/cm$^3$ and a yield strength of 750 MPa. We considered idealized penetrator masses of 100 kg and 500 kg, as outlined in Sections \ref{sec:20m intercept} and \ref{sec:50m intercept}. However, discrepancies arise between idealized and realized penetrator masses due to mesh resolution effects. To minimize these discrepancies, we selected mesh resolutions such that the realized penetrator mass differed from the idealized mass by less than a factor of two.

The simulations presented here are conducted in two stages. Given that the penetrator dimensions are significantly smaller than the rubble-pile asteroid model, a non-uniform mesh is used in the first simulation stage to resolve the early-time dynamics of the hypervelocity impact. In this approach, the resolution increases as the distance to the initial velocity vector of the projectile model decreases. This fine resolution in the area of interest enables high-fidelity modeling of the penetrator and its interactions with the asteroid material. At $t=10$ milliseconds after impact, we transition to the second simulation stage, for which the final state of the first early-time simulation is used as the starting point. The second simulation stage utilizes a uniform mesh with 1 m$^3$ resolution throughout a much larger region in order to extend the simulation to macroscopic timescales. This pseudo-adaptive meshing strategy ensured computational efficiency while maintaining sufficient accuracy for both early and late stages of the simulation.

\subsection{Rubble-pile asteroid model}

We model analogues for 2023 NT1 within the 20--50 m diameter range as heterogeneous rubble piles, consisting of spherical boulders with varying scalar yield strengths embedded within a weak binder material. ``Yield strength" refers to the material's scalar yield strength ($\sigma_y$) at zero pressure, which is independent from tension or compression. The boulders are randomly distributed within the parent model's predefined target volume to achieve a specific fill fraction. Individual boulder sizes are determined by specifying discrete volume fill fractions that align with a target distribution. Within the simulation, the distribution spans boulder diameters from 0.5 m to 8 m, as is shown by the red bar chart in Figure \ref{fig:20m impact sim}. This approximates the power-law distribution of boulder sizes observed during NASA's Double Asteroid Redirection Test (DART) on the surfaces of Didymos and Dimorphos \citep{jewitt_dimorphos_2023, pajola_boulder_2023}. 

We incorporate a porous crush model for both the binder and boulder materials, \textcolor{blue}{based on the p-alpha model developed by \citep{herrmann_constitutive_1969} and further expanded upon by \citep{housen_impacts_2018}.} \textcolor{blue}{We assign porosity values of} 50\% to the binder and 40\% to the boulders, based upon findings by Housen et al. on the impact of porosity in asteroid collisions\ \citep{housen_impacts_2018}. This porosity parameter modifies the material density and defines a pressure threshold, commonly referred to as the ``crush pressure." Below this threshold, any applied pressure is absorbed, crushing the pores rather than damaging the host material. The crush pressures for 40\% boulder porosity and 50\% binder porosity are approximately 100 MPa and 30 MPa, respectively. \textcolor{blue}{For more information on our porosity and material models, see Section 2.2 in our previous work \citep{cohen_asteroid_2024}.}

We vary the initial yield strength of the asteroid material with seven separate Weibull distributions, six of which are used for the boulder distribution, as shown in Figure \ref{fig:weibull} \citep{weibull_statistical_1951}. A seventh distribution is assigned to the binder material, with a weak strength distribution ranging from approximately 1--50 Pa, based on studies of asteroid rotation rates and boulder size distributions \citep{sanchez_strength_2014,jewitt_dimorphos_2023, pajola_boulder_2023}. Weibull distributions are commonly used to model the strength distributions in brittle materials, and the specific Weibull strength distributions used for the boulders approximate the strengths of various common terrestrial materials.

Additionally, we model the bolide's interior structure as having large strength variation, mirroring the complex and heterogeneous structure observed on the surfaces of many rubble-pile asteroids \citep{robin_mechanical_2024}. However, since the interior structure of such asteroids is not yet well understood, this assumption is speculative. Future missions to asteroids are expected to enhance our understanding and enable more accurate models of asteroid interiors.

The weakest boulders have yield strengths ranging from 1--5 MPa to approximate hardened soil, while the strongest boulders range from 250--500 MPa, approximating materials like hardened steel or titanium. These strength values are conservative, likely overestimating the actual strength of boulders in real asteroids. This conservative approach helps to establish practical upper and lower bounds, anchoring our models to testable terrestrial materials, such as granite, for which we have well-defined and reliable equations of state.

For a detailed description covering the disruption simulations, porous crush model, baseline bolide model, yield strength distributions, and other input parameters used for our simulations in ALE3D, please refer to our previous publication by Cohen et al. \citep{cohen_asteroid_2024}.

Given the uncertainty in the diameter of asteroid 2023 NT1, we simulate the interception of both a 20 m and a 50 m diameter case, each with 2.67 g/cm$^3$ average density. While it is estimated that asteroid 2023 NT1 would have had an Earth impact velocity of 15.59 km/s \citep{noauthor_sentry_nodate}, it is within reason to expect that current launch vehicles, such as the SpaceX Falcon 9 (and similar vehicles), could deliver up to 2500 kg of payload mass to the target with characteristic energy C$_{3}$ of $\sim$10 km$^2$/s$^2$, resulting in a closing velocity between the asteroid and the penetrator(s) of $\sim$20 km/s. We use this value for all impact velocities in our hypervelocity impact simulations presented below. 

\subsection{20 m diameter threat with 100 kg penetrator mass} \label{sec:20m intercept}
In our first hypothetical mitigation simulation, we model asteroid 2023 NT1 as a 20 m spherical asteroid using the heterogeneous material model described above. We found that the impact of a cylindrical tungsten penetrator with realized mass 137.4 kg (100 kg idealized mass) at an intercept velocity of 20 km/s delivers a specific impact energy of approximately 1,309 J/kg (defined as the kinetic energy of the penetrator divided by the mass of the target). This specific impact energy is greater than 13 times the catastrophic disruption limit of 100 J/kg (for sphere-on-sphere impacts) described in \citep{jutzi_fragment_2010}. Thus, as expected, the penetrator is highly effective at disrupting the 20 m asteroid into fragments smaller than 5 m. Figure \ref{fig:20m impact sim} shows the results of this simulation at $t$ = 5 seconds after impact. As seen in the histograms, the mean fragment diameter is 1 m, and the mean fragment velocity is 9.3 m/s, which is 1,000 times greater than the gravitational escape velocity of the original asteroid. As a result, the fragments would not be expected to recombine due to gravitational attraction.

\subsection{50 m diameter threat with 500 kg penetrator mass} \label{sec:50m intercept}
By scaling the total idealized penetrator mass from 100 kg to 500 kg, either in the form of a single 500 kg penetrator (referred to as the 1$\times$500 kg case, with a realized mass of 883.7 kg) or in the form of five 100 kg penetrators arranged in a +-shaped array (referred to as the 5$\times$100 kg case, with a realized mass of 146.6 kg per penetrator), we find that we can achieve sufficient disruption of a 50 m analogue of asteroid 2023 NT1. With a kinetic energy of $\sim176.6$ GJ, the specific impact energy delivered in the $1\times500$ kg case is approximately 736.4 J/kg, which is greater than 7 times the 100 J/kg disruption limit mentioned in Section \ref{sec:20m intercept}.

\begin{figure*}
    \centering
    \includegraphics[width=0.9\textwidth]{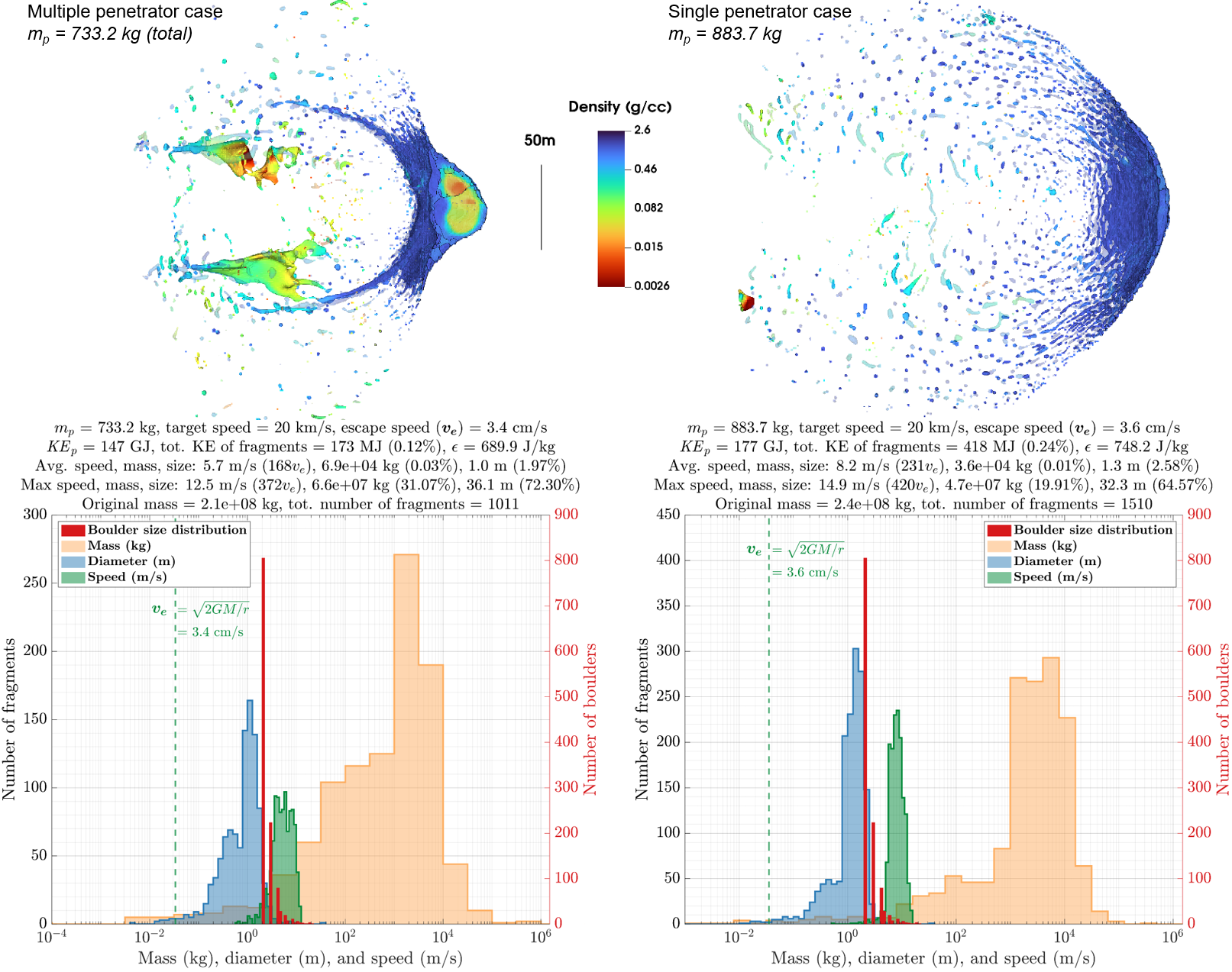}
    \caption{Fragment statistics at $t$ = 10 seconds after impact for the 5$\times$100 kg case (left) and the 1$\times$500 kg case (right), both incident at 20 km/s upon identical 50 m diameter rubble pile asteroid targets. The orange, blue, and green histograms indicate the distributions of fragment masses in kilograms, diameters in meters, and velocity in meters per second, respectively. The effectiveness of disruption can be quantified by comparing the 100 GJ initial kinetic energy of the penetrator(s) to the total kinetic energy of the fragments at a later time. In these cases, we see 0.17\% of the original kinetic energy has been transferred to the fragments in the 5$\times$100 kg case, and 0.42\% has been transferred to the fragments in the 1$\times$500 kg case. It is clear that the 1$\times$500 kg case succeeds in transferring $>$2 times more of the initial kinetic energy of the penetrator than the 5$\times$100 kg case. This figure from \citep{cohen_asteroid_2024} is licensed under \href{https://creativecommons.org/licenses/by-nc-nd/4.0/deed.en}{CC BY-NC-ND 4.0}.}
    \label{fig:50m impact sim}
\end{figure*}

Figure \ref{fig:50m impact sim} compares the two 500 kg cases described above and illustrates how concentrating the penetrator mass into a single penetrator results in more than twice the coupling efficiency between the kinetic energy of the penetrator and the bulk kinetic energy of the fragments after impact. This is likely due to the greater depth of penetration achieved by the larger penetrator, which results in a greater tamping effect on the explosive expansion of the superheated material local to the impact site. However, both cases result in catastrophic disruption of the asteroid and mean fragment velocities which are more than 100 times the gravitational escape velocity of the original asteroid. 

%

As shown in Figure \ref{fig:50m impact sim}, at $t$ = 10 seconds after impact, there are a total of 1,010 fragments ($\geq$1 m) in the 5$\times$100 kg case and 1,509 fragments ($\geq$1 m) in the 1$\times$500 kg case. However, at this time step, both simulations still exhibit a larger, more persistent fragment at the right edge. In both cases, this larger fragment is in the act of disassembling, which is a process that evolves over timescales longer than 10 seconds, as evidenced by its continued deformation. The majority of this persistent fragment is composed of material that has completely failed mechanically. As such, material in the failed state is bound only by frictional cohesion and gravitational binding, both of which are greatly overcome by the fragment velocity imparted by the impact. These factors combined support the hypothesis that the largest remaining fragment will dissociate into fragments of size comparable to the original interior boulder distribution, though likely even smaller, due to the boulder material having already failed.

\section{Ground effects} \label{sec:ground effects}
Following interception and disassembly of the asteroid, its initial exo-atmospheric kinetic energy is distributed into the fragment cloud, which, upon entry with Earth’s atmosphere, results in a series of airburst events. We simulate the ground effects produced by these airburst events for the terminal mode relevant in this paper via computation of the optical pulse model and analytic airburst model for a given scenario.

A full description of the methodology and simulation process for the ground effects models–including parameter definition, computation, and statistical variation–can be found in \citep{bailey_optical_2024}.

\subsection{Optical pulse modeling} \label{sec:optical ground effects}

The conversion of kinetic energy into optical energy is highly dependent on fragment properties, particularly cohesive strength, and is poorly understood in general. We resort to measured optical data to model this conversion, primarily from Department of Defense satellite observations of a small number of relevant bolide sizes of interest to us (typically 1–15 m in diameter) \citep{brown_flux_2002}.

Using an analytical extrapolation from \citep{brown_flux_2002}, we calculate the optical energy flux at burst (in Joules) from exo-atmospheric energy $E_{exo}$ as
\begin{equation} \label{eq.analytical op}
    E_{opt} = \left( \frac{E_{exo}}{8.2508} \right) ^{1.13}
\end{equation}

For the propagation of the optical pulse through the atmosphere, we use a full radiation transfer model to compute the optical power flux from each fragment at each observer. Because the optical propagation is occurring at very close to the speed of light in vacuum and the relevant distance scale from the fragment to the observer are of order of tens to hundreds of km, the light propagation time scale is very short. The optical pulse can then be well approximated as happening nearly simultaneously at all observer points, with the optical pulse arriving very shortly after the fragment burst. 

As a result, we propagate the optical energy flux to each observer using the distance propagation of light, inputting $E_{opt}$ at the time of burst to simulate an instantaneous addition of energy flux \citep{lubin_asteroid_2023,bailey_optical_2024}. We propagate the optical emission from every fragment to the observer to get a total energy flux at the observer.

\subsection{\label{sec:blackbody}Atmospheric attenuation and cooling}
Several factors can greatly affect the observed optical flux, such as the source spectral energy distribution and atmospheric attenuation, which depend on both the source and the complex and time-varying nature of the atmosphere \citep{glasstone_effects_1977}. To calculate the attenuation of the optical signature, we approximate the parent asteroid as a blackbody source due to the wavelength-dependent transmission of the atmosphere \citep{lubin_asteroid_2023}. We model the atmosphere using MODTRAN to perform a full analysis of attenuation which includes the curvature of Earth’s atmosphere and the relationships between nominal atmospheric pressure, temperature, and altitude \citep{lubin_asteroid_2023}. 

Note that we assume an extremely conservative case of no cooling at the observer between fragment optical pulses. However, in a real scenario, fragments arrive on the order of tens of seconds apart (to hundreds of seconds in some extreme cases), which would, in general, be enough time for significant cooling between bursts of incident optical energy. This will be important when judging the effectiveness of this method if the energy exceeds our threshold.

\subsection{Acoustic shock wave modeling} \label{sec:acoustic ground effects}
The acoustic ground effects from an asteroid fragment airburst can be related to and approximated by those of nuclear blasts as discussed by \citep{boslough_updated_2015}. As such, we base our simulations of the acoustic ground effects from mitigation via PI on measurements of equivalent nuclear blasts. To model the time evolution of the shock wave, we use a Friedlander functional form, given by
\begin{equation}\label{time evolution}
    p(t,r)=p_0(r)e^{-t/t_1}(1-t/t_1)
\end{equation}
which describes the shock wave time evolution at a distance $r$ with two free parameters: the peak pressure $p_0$ in Pa at time $t=0$ seconds and the Friedlander positive pulse time scale, or zero crossing time, $t_1$ in seconds \citep{bailey_optical_2024}. Note that the time $t=0$ seconds is the time at which the shock wave first arrives at the observer, not to be confused with the time at which the fragment bursts.

Letting $\epsilon$ denote the fraction of a 1 kt yield that goes into the shock wave (typically 0.5), we achieve $E_{nuc}=E_{ast-kt}/\epsilon$, where $E_{ast-kt}$ is the asteroid airburst shock wave energy and $E_{nuc}$ is the equivalent energy of a nuclear weapon \citep{lubin_asteroid_2023}. The peak pressure at a distance $r$ is calculated from the equivalent energy of a nuclear weapon as
\begin{equation}\label{peak pressure}
    p(r)=p_n{[rE_{nuc}^{1/3}]}^{\alpha_n}+p_f{[rE_{nuc}^{1/3}]}^{\alpha_f}
\end{equation}
where $p_n=3.11\times10^{11}$ Pa is the pressure for a 1 kt standard weapon yield in the near field, $\alpha_n=-2.95$ is the power law index for the near field, $p_f=1.80\times10^{7}$ Pa is the pressure for a 1 kt standard weapon yield in the far field, and $\alpha_f=-1.13$ is the power law index for the far field \citep{lubin_asteroid_2023}.

We simulate the airburst produced by each fragment as it enters Earth's atmosphere using equations (\ref{time evolution}) and (\ref{peak pressure}). The model considers any interference between interacting shock waves, summing them to simulate the acoustic caustics.

\subsection{\label{sec:caustics}Shock wave caustics}
As a fragment airbursts, the emitted shock wave can be thought of as an expanding sphere whose intersection with the ground plane forms a circle. Acoustic caustics form when shock waves from multiple fragment airbursts constructively interfere in the ground plane. Our model takes into account such interference to simulate the acoustic caustics that form as shock waves interact \citep{lubin_asteroid_2023}. Areas of constructive interference experience higher over-pressures, and thus they must be taken into account in order to design mitigation scenarios with acceptably low pressure values.

\subsection{Terminal mitigation scenarios for 2023 NT1} \label{sec:ground effects scenarios}
Given the large uncertainty in 2023 NT1's physical characteristics, we investigated more than 100 threat case simulations that vary both the body's physical parameters (e.g., size, average density, and incident entry angle relative to Earth's horizon) and mitigation parameters (e.g., number of fragments and intercept times). A summary of all of the cases considered during the parameter sweep can be found in Tables \ref{table: frag} and \ref{table: unfrag}. Each scenario is designed to keep the ground effects (as experienced by an arbitrary observer on Earth’s surface) below the damage thresholds outlined in Section \ref{sec:mitigation} to minimize ground damage. As such, we utilize the results of our hypervelocity impact simulations as proxies for mitigation parameters. 

Following the histogram distributions for fragment size in Figures \ref{fig:20m impact sim} and \ref{fig:50m impact sim}, we vary the total number of fragments to keep the average fragment size for each case at or below $\sim$4 m in diameter (average fragment size over all mitigated cases = 4.00 m, Table \ref{table: frag}). Note that several individual cases have average fragment sizes larger than 4 m; these represent extreme cases with a conservative number of fragments relative to the threat size. Such scenarios are included for comparison to other threat sizes; e.g., a 60 m asteroid with an average density of 2.6 g/cm$^3$ and entry angle of 45\textdegree\ which is broken into 1000 fragments with a one-day intercept (case 92, Table \ref{table: frag}) is included for comparison with 26–58 m asteroids with the same physical parameters and mitigation scenario.

We introduce statistical variations in the fragmentation process to simulate the uncertainty in a real scenario by defining normal distributions for several fragment parameters, including diameter (fragment size), disruption velocity (the asymptotic velocity at which fragments move away from the fragment cloud’s center of mass after having been decelerated by gravitational attraction of the asteroid), density, yield strength, and slant distance from the airburst. A thorough explanation of how we introduce statistical variation into the model, as well as several examples, can be found in \citep{bailey_optical_2024}. All scenarios assume a spherical target body. Mitigated scenarios assume an average fragment disruption velocity of 1 m/s.

We also model the ground effects of unmitigated (i.e., unfragmented) scenarios as a comparison (Table \ref{table: unfrag}). All unmitigated cases assume a spherical body with an entry angle of 45\textdegree\ relative to Earth's horizon. As the simulations show in the following sections, the damage to life and infrastructure caused by an unmitigated airburst is far greater than that of an equal mitigated case, the latter of which results in ground effects below estimated damage thresholds.

All scenarios, mitigated and unmitigated, assume an impact velocity (the velocity at atmospheric entry if an impact had occurred) of 15.59 km/s as estimated by JPL Sentry \citep{noauthor_sentry_nodate}, as mentioned in Section \ref{sec:2023 NT1 params}. Thus, the scenarios outlined here simulate hypothetical outcomes if 2023 NT1 had hit Earth on its closest approach date of July 13, 2023, instead of narrowly missing.

\startlongtable
\begin{deluxetable*}{cccccccccccc}
\tablecaption{Summary of mitigated (fragmented) threat scenarios and estimated ground effects. All scenarios assume a spherical parent asteroid with an impact velocity of 15.59 km/s and average fragment disruption of 1 m/s. \textit{D} indicates parent asteroid diameter; $\rho$ indicates average parent asteroid density; intercept time is defined as the length of time prior to ground impact that the asteroid is intercepted.}
\tabletypesize{\scriptsize}
\tablenum{1}

  \tablehead{\colhead{Case} & \colhead{D} & \colhead{$\rho$} & \colhead{Entry} & \colhead{No.} & \colhead{Avg.} & \colhead{Inter-} & \colhead{Unbroken} & \colhead{1\% opt.} & \colhead{Weighted avg.} & \colhead{1\% ac.} & \colhead{Weighted avg.} \\
 \colhead{no.} & \colhead{} & \colhead{} & \colhead{angle} & \colhead{frag.} & \colhead{frag. size} & \colhead{cept} & \colhead{exo-atm.} & \colhead{CDF value} & \colhead{opt. energy} & \colhead{CDF value} & \colhead{pressure} \\
 \colhead{} & \colhead{(m)} & \colhead{(kg/m$^3$)} & \colhead{(\textdegree)} & \colhead{} & \colhead{(m)} & \colhead{time} & \colhead{energy (Mt)} & \colhead{(J/m$^2$)} & \colhead{(J/m$^2$)} & \colhead{(Pa)} & \colhead{(Pa)} }
\startdata
1 & 26 & 2600 & 20 & 1000 & 2.60 & 1 d & 6.95E-01 & 1.67E+04 & 6.86E+03 & 3.96E+02 & 1.59E+02 \\  
        2 & 26 & 2600 & 45 & 1000 & 2.60 & 1 d & 6.95E-01 & 1.11E+03 & 4.06E+02 & 3.49E+02 & 1.41E+02 \\  
        3 & 26 & 2600 & 45 & 1000 & 2.60 & 10 d & 6.95E-01 & 4.74E+01 & 1.11E+01 & 1.44E+02 & 4.33E+01 \\  
        4 & 26 & 2600 & 45 & 1000 & 2.60 & 12 hr & 6.95E-01 & 3.13E+03 & 1.34E+03 & 5.49E+02 & 2.50E+02 \\  
        5 & 26 & 2600 & 70 & 1000 & 2.60 & 1 d & 6.95E-01 & 2.05E+04 & 8.43E+03 & 5.04E+02 & 1.95E+02 \\  
        6 & 26 & 2600 & 90 & 1000 & 2.60 & 1 d & 6.95E-01 & 1.84E+03 & 6.73E+02 & 5.12E+02 & 1.99E+02 \\  
        7 & 26 & 2600 & 45 & 4000 & 1.64 & 1 d & 6.95E-01 & 7.92E+02 & 2.82E+02 & 3.83E+02 & 1.51E+02 \\  
        8 & 26 & 2600 & 45 & 4000 & 1.64 & 10 d & 6.95E-01 & 1.82E+01 & 4.84E+00 & 1.08E+02 & 3.41E+01 \\  
        9 & 26 & 2600 & 45 & 4000 & 1.64 & 12 hr & 6.95E-01 & 2.44E+03 & 9.05E+02 & 6.83E+02 & 3.15E+02 \\  
        10 & 30 & 1400 & 45 & 1000 & 3.00 & 1 d & 5.75E-01 & 7.31E+02 & 2.39E+02 & 2.80E+02 & 1.15E+02 \\  
        11 & 30 & 2600 & 20 & 1000 & 3.00 & 1 d & 1.07E+00 & 1.69E+03 & 6.39E+02 & 3.66E+02 & 1.45E+02 \\  
        12 & 30 & 2600 & 45 & 1000 & 3.00 & 1 d & 1.07E+00 & 2.04E+03 & 7.75E+02 & 4.31E+02 & 1.70E+02 \\  
        13 & 30 & 2600 & 45 & 1000 & 3.00 & 1 hr & 1.07E+00 & 1.36E+04 & 8.32E+03 & 1.64E+03 & 8.71E+02 \\  
        14 & 30 & 2600 & 70 & 1000 & 3.00 & 1 d & 1.07E+00 & 2.29E+03 & 8.58E+02 & 4.67E+02 & 1.88E+02 \\  
        15 & 30 & 2600 & 90 & 1000 & 3.00 & 1 d & 1.07E+00 & 2.38E+03 & 8.74E+02 & 4.79E+02 & 1.92E+02 \\  
        16 & 30 & 4800 & 45 & 1000 & 3.00 & 1 d & 1.97E+00 & 6.71E+03 & 2.53E+03 & 8.02E+02 & 2.98E+02 \\  
        17 & 30 & 6000 & 45 & 2000 & 2.38 & 1 d & 2.46E+00 & 1.14E+04 & 4.18E+03 & 1.15E+03 & 4.04E+02 \\  
        18 & 34 & 1400 & 45 & 1000 & 3.40 & 1 d & 8.37E-01 & 1.19E+03 & 4.46E+02 & 3.35E+02 & 1.40E+02 \\  
        19 & 34 & 2600 & 20 & 1000 & 3.40 & 1 d & 1.55E+00 & 2.97E+03 & 1.13E+03 & 4.32E+02 & 1.73E+02 \\  
        20 & 34 & 2600 & 45 & 1000 & 3.40 & 1 d & 1.55E+00 & 3.47E+03 & 1.30E+03 & 5.09E+02 & 1.99E+02 \\  
        21 & 34 & 2600 & 45 & 1000 & 3.40 & 10 d & 1.55E+00 & 1.42E+02 & 3.42E+01 & 2.15E+02 & 6.35E+01 \\  
        22 & 34 & 2600 & 45 & 1000 & 3.40 & 12 hr & 1.55E+00 & 8.98E+03 & 4.22E+03 & 8.03E+02 & 3.61E+02 \\  
        23 & 34 & 2600 & 70 & 1000 & 3.40 & 1 d & 1.55E+00 & 3.78E+03 & 1.42E+03 & 5.60E+02 & 2.16E+02 \\  
        24 & 34 & 2600 & 90 & 1000 & 3.40 & 1 d & 1.55E+00 & 3.80E+03 & 1.48E+03 & 5.68E+02 & 2.22E+02 \\  
        25 & 34 & 2600 & 45 & 4000 & 2.14 & 1 d & 1.55E+00 & 1.95E+03 & 7.04E+02 & 4.67E+02 & 1.85E+02 \\  
        26 & 34 & 2600 & 45 & 4000 & 2.14 & 10 d & 1.55E+00 & 5.13E+01 & 1.37E+01 & 1.37E+02 & 4.41E+01 \\  
        27 & 34 & 2600 & 45 & 4000 & 2.14 & 12 hr & 1.55E+00 & 5.56E+03 & 2.27E+03 & 8.11E+02 & 3.70E+02 \\  
        28 & 34 & 4000 & 45 & 1000 & 3.40 & 1 d & 2.39E+00 & 7.77E+03 & 2.99E+03 & 7.76E+02 & 2.83E+02 \\  
        29 & 34 & 6000 & 45 & 1000 & 3.40 & 1 d & 3.59E+00 & 2.12E+04 & 7.78E+03 & 1.42E+03 & 5.10E+02 \\  
        30 & 35 & 1800 & 45 & 1000 & 3.50 & 1 d & 1.17E+00 & 1.94E+03 & 7.67E+02 & 4.04E+02 & 1.64E+02 \\  
        31 & 35 & 2600 & 20 & 1000 & 3.50 & 1 d & 1.70E+00 & 3.42E+03 & 1.28E+03 & 4.55E+02 & 1.82E+02 \\  
        32 & 35 & 2600 & 45 & 1000 & 3.50 & 1 d & 1.70E+00 & 3.42E+03 & 1.28E+03 & 5.35E+02 & 2.10E+02 \\  
        33 & 35 & 2600 & 70 & 1000 & 3.50 & 1 d & 1.70E+00 & 4.42E+03 & 1.66E+03 & 5.93E+02 & 2.29E+02 \\  
        34 & 35 & 2600 & 90 & 1000 & 3.50 & 1 d & 1.70E+00 & 4.70E+03 & 1.72E+03 & 6.09E+02 & 2.38E+02 \\  
        35 & 35 & 4000 & 45 & 1000 & 3.50 & 1 d & 2.61E+00 & 9.40E+03 & 3.43E+03 & 8.22E+02 & 3.11E+02 \\  
        36 & 35 & 6000 & 45 & 2000 & 2.78 & 1 d & 3.91E+00 & 1.94E+04 & 7.08E+03 & 1.37E+03 & 4.83E+02 \\  
        37 & 40 & 2600 & 20 & 1000 & 4.00 & 1 d & 2.53E+00 & 5.28E+03 & 2.06E+03 & 5.33E+02 & 2.11E+02 \\  
        38 & 40 & 2600 & 45 & 1000 & 4.00 & 1 d & 2.53E+00 & 5.28E+03 & 2.06E+03 & 6.69E+02 & 2.48E+02 \\  
        39 & 40 & 2600 & 70 & 1000 & 4.00 & 1 d & 2.53E+00 & 7.52E+03 & 2.86E+03 & 7.25E+02 & 2.77E+02 \\  
        40 & 40 & 2600 & 90 & 1000 & 4.00 & 1 d & 2.53E+00 & 7.49E+03 & 2.86E+03 & 7.33E+02 & 2.84E+02 \\  
        41 & 40 & 2700 & 45 & 1000 & 4.00 & 1 d & 2.63E+00 & 6.62E+03 & 2.64E+03 & 6.62E+02 & 2.65E+02 \\  
        42 & 40 & 3400 & 45 & 1000 & 4.00 & 1 d & 3.31E+00 & 1.15E+04 & 4.16E+03 & 8.31E+02 & 3.22E+02 \\  
        43 & 40 & 6000 & 45 & 2000 & 3.17 & 1 d & 5.84E+00 & 3.16E+04 & 1.17E+04 & 1.63E+03 & 5.79E+02 \\  
        44 & 45 & 2600 & 20 & 1000 & 4.50 & 1 d & 3.60E+00 & 7.99E+03 & 3.18E+03 & 6.14E+02 & 2.44E+02 \\  
        45 & 45 & 2600 & 45 & 1000 & 4.50 & 1 d & 3.60E+00 & 7.99E+03 & 3.18E+03 & 8.04E+02 & 3.11E+02 \\  
        46 & 45 & 2600 & 70 & 1000 & 4.50 & 1 d & 3.60E+00 & 1.17E+04 & 4.50E+03 & 8.64E+02 & 3.28E+02 \\  
        47 & 45 & 2600 & 90 & 1000 & 4.50 & 1 d & 3.60E+00 & 1.13E+04 & 4.44E+03 & 8.70E+02 & 3.39E+02 \\  
        48 & 45 & 2800 & 45 & 1000 & 4.50 & 1 d & 3.88E+00 & 1.24E+04 & 4.73E+03 & 8.31E+02 & 3.13E+02 \\  
        49 & 45 & 4000 & 45 & 1000 & 4.50 & 1 d & 5.54E+00 & 2.53E+04 & 9.05E+03 & 1.18E+03 & 4.36E+02 \\  
        50 & 45 & 4000 & 45 & 1000 & 4.50 & 12 hr & 5.54E+00 & 5.71E+04 & 2.68E+04 & 1.87E+03 & 8.18E+02 \\  
        51 & 45 & 4000 & 45 & 1000 & 4.50 & 1 hr & 5.54E+00 & 1.84E+05 & 9.50E+04 & 5.93E+03 & 3.10E+03 \\  
        52 & 45 & 6000 & 45 & 4000 & 2.83 & 1 d & 8.31E+00 & 4.42E+04 & 1.63E+04 & 1.91E+03 & 6.57E+02 \\  
        53 & 50 & 2600 & 20 & 1000 & 5.00 & 1 d & 4.94E+00 & 1.39E+04 & 5.36E+03 & 7.57E+02 & 3.04E+02 \\  
        54 & 50 & 2600 & 45 & 1000 & 5.00 & 1 d & 4.94E+00 & 1.39E+04 & 5.36E+03 & 9.15E+02 & 3.54E+02 \\  
        55 & 50 & 2600 & 70 & 1000 & 5.00 & 1 d & 4.94E+00 & 1.39E+04 & 5.36E+03 & 1.02E+03 & 3.90E+02 \\  
        56 & 50 & 2600 & 90 & 1000 & 5.00 & 1 d & 4.94E+00 & 1.39E+04 & 5.36E+03 & 1.09E+03 & 3.99E+02 \\  
        57 & 50 & 3100 & 45 & 1000 & 5.00 & 1 d & 5.89E+00 & 2.26E+04 & 8.59E+03 & 1.08E+03 & 3.96E+02 \\  
        58 & 50 & 4500 & 45 & 2000 & 5.00 & 1 d & 8.55E+00 & 3.52E+04 & 1.34E+04 & 1.44E+03 & 5.16E+02 \\  
        59 & 50 & 6000 & 45 & 4000 & 3.15 & 1 d & 1.14E+01 & 6.10E+04 & 2.30E+04 & 2.14E+03 & 7.37E+02 \\  
        60 & 50 & 6000 & 45 & 4000 & 3.15 & 2 d & 1.14E+01 & 2.15E+04 & 7.10E+03 & 1.31E+03 & 3.85E+02 \\  
        61 & 55 & 2600 & 20 & 1000 & 5.50 & 1 d & 6.58E+00 & 1.72E+04 & 7.23E+03 & 8.46E+02 & 3.42E+02 \\  
        62 & 55 & 2600 & 45 & 1000 & 5.50 & 1 d & 6.58E+00 & 2.22E+04 & 8.81E+03 & 1.06E+03 & 4.04E+02 \\  
        63 & 55 & 2600 & 70 & 1000 & 5.50 & 1 d & 6.58E+00 & 2.68E+04 & 1.03E+04 & 1.21E+03 & 4.41E+02 \\  
        64 & 55 & 2600 & 90 & 1000 & 5.50 & 1 d & 6.58E+00 & 2.81E+04 & 1.08E+04 & 1.26E+03 & 4.64E+02 \\  
        65 & 55 & 2600 & 20 & 2000 & 5.50 & 1 d & 6.58E+00 & 1.84E+04 & 7.26E+03 & 1.61E+03 & 3.80E+02 \\  
        66 & 55 & 2600 & 45 & 2000 & 5.50 & 1 d & 6.58E+00 & 1.84E+04 & 7.26E+03 & 9.57E+02 & 3.68E+02 \\  
        67 & 55 & 2600 & 70 & 2000 & 5.50 & 1 d & 6.58E+00 & 1.84E+04 & 7.26E+03 & 2.13E+03 & 4.80E+02 \\  
        68 & 55 & 2600 & 90 & 2000 & 5.50 & 1 d & 6.58E+00 & 1.84E+04 & 7.26E+03 & 2.18E+03 & 4.92E+02 \\  
        69 & 55 & 3800 & 45 & 2000 & 5.50 & 1 d & 9.61E+00 & 3.56E+04 & 1.37E+04 & 1.35E+03 & 4.88E+02 \\  
        70 & 55 & 6000 & 45 & 4000 & 3.46 & 1 d & 1.52E+01 & 8.77E+04 & 3.31E+04 & 2.46E+03 & 8.40E+02 \\  
        71 & 55 & 6000 & 45 & 4000 & 3.46 & 2 d & 1.52E+01 & 2.82E+04 & 9.51E+03 & 1.45E+03 & 4.31E+02 \\  
        72 & 55 & 6000 & 45 & 6000 & 3.03 & 1 d & 1.52E+01 & 7.94E+04 & 3.00E+04 & 2.41E+03 & 8.27E+02 \\  
        73 & 55 & 6000 & 45 & 6000 & 3.03 & 2 d & 1.52E+01 & 2.64E+04 & 9.12E+03 & 1.43E+03 & 4.22E+02 \\  
        74 & 58 & 1400 & 45 & 1000 & 5.80 & 1 d & 4.15E+00 & 1.20E+04 & 4.72E+03 & 798.3654 & 3.13E+02 \\  
        75 & 58 & 2600 & 20 & 1000 & 5.80 & 1 d & 7.71E+00 & 2.13E+04 & 8.60E+03 & 902.1732 & 3.52E+02 \\  
        76 & 58 & 2600 & 45 & 1000 & 5.80 & 1 d & 7.71E+00 & 2.87E+04 & 1.12E+04 & 1171.851 & 4.38E+02 \\  
        77 & 58 & 2600 & 45 & 1000 & 5.80 & 10 d & 7.71E+00 & 1.19E+03 & 2.80E+02 & 452.8814 & 1.27E+02 \\  
        78 & 58 & 2600 & 45 & 1000 & 5.80 & 12 hr & 7.71E+00 & 7.51E+04 & 3.58E+04 & 1910.354 & 8.32E+02 \\  
        79 & 58 & 2600 & 70 & 1000 & 5.80 & 1 d & 7.71E+00 & 3.26E+04 & 1.29E+04 & 1330.377 & 4.75E+02 \\  
        80 & 58 & 2600 & 90 & 1000 & 5.80 & 1 d & 7.71E+00 & 3.47E+04 & 1.25E+04 & 1343.861 & 4.99E+02 \\  
        81 & 58 & 2600 & 45 & 4000 & 3.65 & 1 d & 7.71E+00 & 1.81E+04 & 7.18E+03 & 965.7292 & 3.75E+02 \\  
        82 & 58 & 2600 & 45 & 4000 & 3.65 & 10 d & 7.71E+00 & 4.61E+02 & 1.37E+02 & 280.8356 & 8.74E+01 \\  
        83 & 58 & 2600 & 45 & 4000 & 3.65 & 12 hr & 7.71E+00 & 4.60E+04 & 2.20E+04 & 1675.361 & 7.63E+02 \\  
        84 & 58 & 4000 & 45 & 1000 & 5.80 & 1 d & 1.19E+01 & 6.65E+04 & 2.42E+04 & 1767.778 & 6.09E+02 \\  
        85 & 58 & 6000 & 45 & 1000 & 5.80 & 1 d & 1.78E+01 & 1.46E+05 & 5.40E+04 & 2936.818 & 1.01E+03 \\  
        86 & 58 & 6000 & 45 & 4000 & 5.80 & 1 d & 1.78E+01 & 1.05E+05 & 3.95E+04 & 2634.387 & 8.92E+02 \\  
        87 & 58 & 6000 & 45 & 4000 & 5.80 & 2 d & 1.78E+01 & 3.50E+04 & 1.17E+04 & 1571.214 & 4.62E+02 \\  
        88 & 60 & 1400 & 45 & 2000 & 4.37 & 1 d & 4.60E+00 & 9.86E+03 & 3.96E+03 & 7.24E+02 & 2.91E+02 \\  
        89 & 60 & 1400 & 45 & 4000 & 3.78 & 2 d & 4.60E+00 & 2.68E+03 & 9.13E+02 & 4.15E+02 & 1.27E+02 \\  
        90 & 60 & 2600 & 20 & 1000 & 6.00 & 1 d & 8.54E+00 & 2.57E+04 & 1.07E+04 & 9.80E+02 & 3.89E+02 \\  
        91 & 60 & 2600 & 20 & 4000 & 3.78 & 2 d & 8.54E+00 & 5.90E+03 & 2.08E+03 & 5.18E+02 & 1.58E+02 \\  
        92 & 60 & 2600 & 45 & 1000 & 6.00 & 1 d & 8.54E+00 & 3.11E+04 & 1.22E+04 & 1.20E+03 & 4.68E+02 \\  
        93 & 60 & 2600 & 45 & 4000 & 3.78 & 2 d & 8.54E+00 & 7.36E+03 & 2.52E+03 & 6.24E+02 & 1.89E+02 \\  
        94 & 60 & 2600 & 70 & 1000 & 6.00 & 1 d & 8.54E+00 & 3.38E+04 & 1.35E+04 & 1.35E+03 & 5.02E+02 \\  
        95 & 60 & 2600 & 70 & 4000 & 3.78 & 2 d & 8.54E+00 & 5.90E+03 & 2.08E+03 & 6.80E+02 & 2.06E+02 \\  
        96 & 60 & 2600 & 90 & 1000 & 6.00 & 1 d & 8.54E+00 & 3.76E+04 & 1.48E+04 & 1.42E+03 & 5.03E+02 \\  
        97 & 60 & 2600 & 90 & 4000 & 3.78 & 2 d & 8.54E+00 & 5.90E+03 & 2.08E+03 & 6.99E+02 & 2.11E+02 \\  
        98 & 60 & 3200 & 45 & 4000 & 3.78 & 2 d & 1.05E+01 & 1.03E+04 & 3.54E+03 & 7.35E+02 & 2.22E+02 \\  
        99 & 60 & 4500 & 45 & 4000 & 3.78 & 2 d & 1.48E+01 & 2.04E+04 & 6.98E+03 & 1.07E+03 & 3.19E+02 \\  
        100 & 60 & 4500 & 45 & 6000 & 3.30 & 2 d & 1.48E+01 & 1.79E+04 & 6.19E+03 & 1.03E+03 & 3.08E+02 \\  
        101 & 60 & 5200 & 45 & 4000 & 3.78 & 2 d & 1.71E+01 & 2.77E+04 & 9.53E+03 & 1.31E+03 & 3.90E+02 \\  
        102 & 60 & 5200 & 45 & 6000 & 3.30 & 2 d & 1.71E+01 & 2.42E+04 & 8.44E+03 & 1.25E+03 & 3.71E+02 \\  
        103 & 60 & 6000 & 45 & 6000 & 3.30 & 1 d & 1.97E+01 & 1.10E+05 & 4.15E+04 & 2.75E+03 & 9.41E+02 \\  
        104 & 60 & 6000 & 45 & 6000 & 3.30 & 2 d & 1.97E+01 & 3.55E+04 & 1.21E+04 & 1.57E+03 & 4.68E+02 \\  
        105 & 60 & 7000 & 45 & 4000 & 3.78 & 2 d & 2.30E+01 & 6.15E+04 & 2.05E+04 & 2.23E+03 & 6.49E+02 \\  
        106 & 60 & 7000 & 45 & 4000 & 3.78 & 5 d & 2.30E+01 & 1.25E+04 & 3.71E+03 & 1.45E+03 & 4.27E+02 \\ 
\enddata



\label{table: frag}
\end{deluxetable*}

\startlongtable
\begin{deluxetable*}{cccccccccc}




\tablecaption{Summary of unmitigated (unfragmented) 2023 NT1 threat scenarios and estimated ground effects. All scenarios assume a spherical parent asteroid with an impact velocity of 15.59 km/s and entry angle of 45\textdegree. \textit{D} indicates asteroid diameter; $\rho$ indicates average asteroid density. Note that the maximum optical energy and maximum pressure values are experienced at the lowest probability level of the simulation, which may vary between cases.}

\tabletypesize{\scriptsize}
\tablenum{2}

\tablehead{\colhead{Case} & \colhead{D} & \colhead{$\rho$} & \colhead{Unbroken} & \colhead{1\% opt.} & \colhead{Weighted avg.} & \colhead{Max. opt.} & \colhead{1\% ac.} & \colhead{Weighted avg.} & \colhead{Max.} \\ 
\colhead{no.} & \colhead{} & \colhead{} & \colhead{exo-atm.} & \colhead{CDF value} & \colhead{opt. energy} & \colhead{energy} & \colhead{CDF value} & \colhead{pressure} & \colhead{pressure} \\
\colhead{} & \colhead{(m)} & \colhead{(kg/m$^3$)} & \colhead{energy (Mt)} & \colhead{(J/m$^2$)} & \colhead{(J/m$^2$)} & \colhead{(J/m$^2$)} & \colhead{(Pa)} & \colhead{(Pa)} & \colhead{(Pa)} } 

\startdata
 107 & 26 & 2600 & 6.95E-01 & 9.42E+04 & 2.26E+04 & 1.05E+05 & 2.91E+03 & 1.33E+03 & 3.03E+03 \\  
        108 & 30 & 2600 & 1.07E+00 & 1.90E+05 & 4.39E+04 & 2.18E+05 & 3.85E+03 & 1.64E+03 & 4.04E+03 \\  
        109 & 34 & 2600 & 1.55E+00 & 3.57E+05 & 8.02E+04 & 4.23E+05 & 4.98E+03 & 1.27E+03 & 5.30E+03 \\  
        110 & 34 & 4000 & 2.39E+00 & 1.03E+06 & 2.21E+05 & 1.39E+06 & 8.15E+03 & 1.76E+03 & 9.26E+03 \\  
        111 & 34 & 6000 & 3.59E+00 & 3.80E+06 & 9.92E+05 & 9.02E+06 & 1.85E+04 & 3.24E+03 & 3.18E+04 \\  
        112 & 35 & 2600 & 1.70E+00 & 4.12E+05 & 9.25E+04 & 4.95E+05 & 5.30E+03 & 1.33E+03 & 5.66E+03 \\  
        113 & 40 & 2600 & 2.53E+00 & 8.29E+05 & 1.81E+05 & 1.05E+06 & 7.17E+03 & 1.67E+03 & 7.86E+03 \\  
        114 & 45 & 2600 & 3.60E+00 & 1.56E+06 & 3.38E+05 & 2.10E+06 & 9.65E+03 & 2.08E+03 & 1.10E+04 \\  
        115 & 45 & 4000 & 5.54E+00 & 4.73E+06 & 1.11E+06 & 8.73E+06 & 1.85E+04 & 3.30E+03 & 2.65E+04 \\  
        116 & 50 & 2600 & 4.94E+00 & 2.83E+06 & 6.10E+05 & 4.08E+06 & 1.30E+04 & 2.58E+03 & 1.56E+04 \\  
        117 & 50 & 4500 & 8.55E+00 & 1.20E+07 & 3.56E+06 & 3.75E+07 & 3.65E+04 & 6.99E+03 & 9.18E+04 \\  
        118 & 55 & 2600 & 6.58E+00 & 4.78E+06 & 1.08E+06 & 7.81E+06 & 1.77E+04 & 3.26E+03 & 2.33E+04 \\  
        119 & 58 & 2600 & 7.71E+00 & 6.54E+06 & 1.50E+06 & 1.14E+07 & 2.14E+04 & 3.80E+03 & 3.02E+04 \\  
        120 & 58 & 4000 & 1.19E+01 & 1.99E+07 & 6.66E+06 & 8.28E+07 & 5.32E+04 & 1.32E+04 & 2.00E+05 \\  
        121 & 58 & 6000 & 1.78E+01 & 6.45E+07 & 4.11E+07 & 2.15E+09 & 1.44E+05 & 1.89E+06 & 1.44E+07 \\  
        122 & 60 & 2600 & 8.54E+00 & 7.88E+06 & 1.87E+06 & 1.48E+07 & 2.44E+04 & 4.26E+03 & 3.66E+04 \\  
        123 & 60 & 5200 & 1.71E+01 & 1.94E+08 & 1.18E+08 & 9.69E+09 & 5.44E+05 & 2.00E+07 & 1.36E+08 \\  
\enddata



\label{table: unfrag}
\end{deluxetable*}

\subsection{Simulation results} \label{sec:ground effects results}
\begin{figure*}[!ht]
    \centering
    \includegraphics[width=0.9\textwidth]{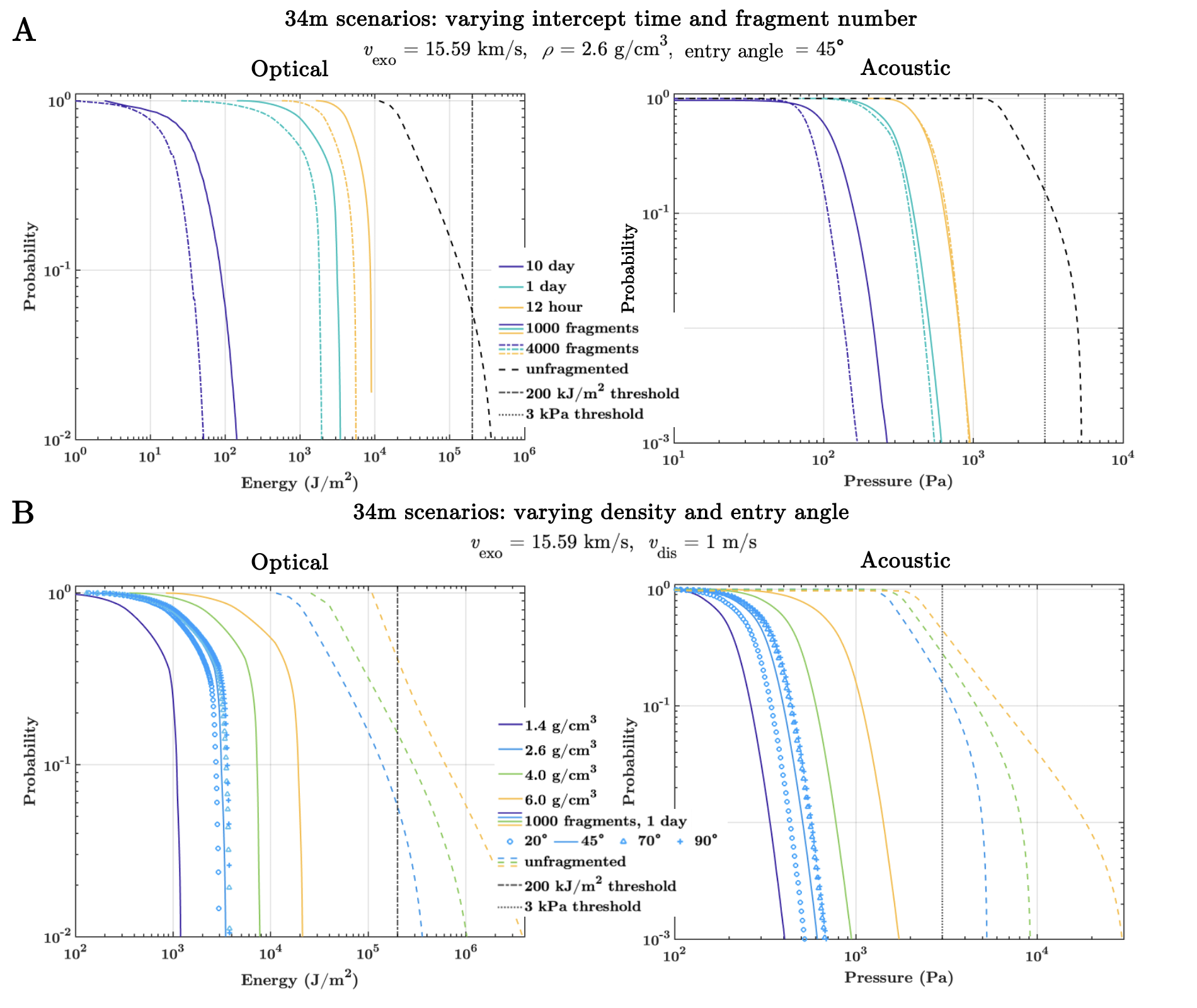}
    \caption{CDFs of the optical (left) and acoustic (right) ground effects of a variety of mitigation scenarios with a 34 m spherical parent asteroid. All scenarios assume an exo-atmospheric parent asteroid velocity (v$_{exo}$) of 15.59 km/s; mitigated scenarios assume an average fragment disruption velocity (v$_{dis}$) of 1 m/s. The dash-dotted black line (left) and dotted black line (right) mark the optical and acoustic damage thresholds of 200 kJ/m$^2$ and 3 kPa, respectively. (A) Varying intercept time and fragment number. All scenarios assume an average density ($\rho$) of 2.6 g/cm$^3$ and entry angle of 45\textdegree\ relative to Earth’s horizon. Legend dictates intercept time prior to impact, number of fragments, and scenario type (fragmented versus unfragmented). (B) Varying density and entry angle. All mitigated scenarios disassemble the parent asteroid into 1000 fragments with a one-day intercept prior to impact. All unmitigated scenarios (dashed lines) assume an entry angle of 45\textdegree\ relative to Earth’s horizon. Unbroken exo-atmospheric energies range from 0.84–3.59 Mt. Legend dictates average density, entry angle relative to Earth's horizon, and scenario type.}
    \label{fig:34m cdf}
\end{figure*}

For our simulations of 2023 NT1, we found that the ground effects of airbursts resulting from mitigation via PI can be kept below their respective damage thresholds (Section \ref{sec:ground effects thresholds}) and are significantly lower than their unmitigated counterparts, when designing scenarios appropriate for the threat.  The magnitude of the effects for any given case is dependent upon the physical parameters of the threat, namely diameter and density,  and specific mitigation parameters, such as the number of resulting fragments and intercept time relative to ground impact. In general, increasing the number of fragments and/or intercept time will decrease the magnitude of the ground effects \citep{lubin_asteroid_2023}.  However, it should be noted that exceptions exist when considering scenarios where target intercepts occur at less than one day prior to impact (Figure \ref{fig:34m cdf}). 

To illustrate the magnitude of the ground effects, we plot the results of our simulations in the form of cumulative distribution functions (CDF) of both optical energy flux and acoustic over-pressure in the ground plane. We use the CDF of a particular threat scenario to determine whether that scenario results in ground effects of acceptably low magnitude. In addition to the maximum value, a useful metric is the 1\% value of the CDF for a particular threat scenario, which is the magnitude at which 1\% of ground locations resolved in the simulation experience optical energy flux or acoustic over-pressure above that value. We refer to this value henceforth as the 1\% CDF value. 

Simulations suggest that PI could effectively mitigate 2023 NT1 throughout a wide range of threat scenarios with very short warning times, from hours to days depending on the object's physical characteristics. Assuming a worst-case scenario of a 60 m diameter iron-nickel asteroid with an average density of 7 g/cm$^3$, we find that disruption of the body into 4000 fragments with an intercept two days prior to impact would be sufficient to keep the vast majority of ground effects below their respective damage thresholds, as indicated by the 1\% CDF values of 61.5 kJ/m$^2$ for optical energy flux and 2.2 kPa for shock wave over-pressure (Table \ref{table: frag}). Further, by increasing the intercept time to five days for the same case, the optical and acoustic ground effect magnitudes are reduced, with 1\% CDF values of 12.5 kJ/m$^2$ and 1.5 kPa, respectively (Table \ref{table: frag}). 

\begin{figure*}
    \centering
    \includegraphics[width=0.725\textwidth]{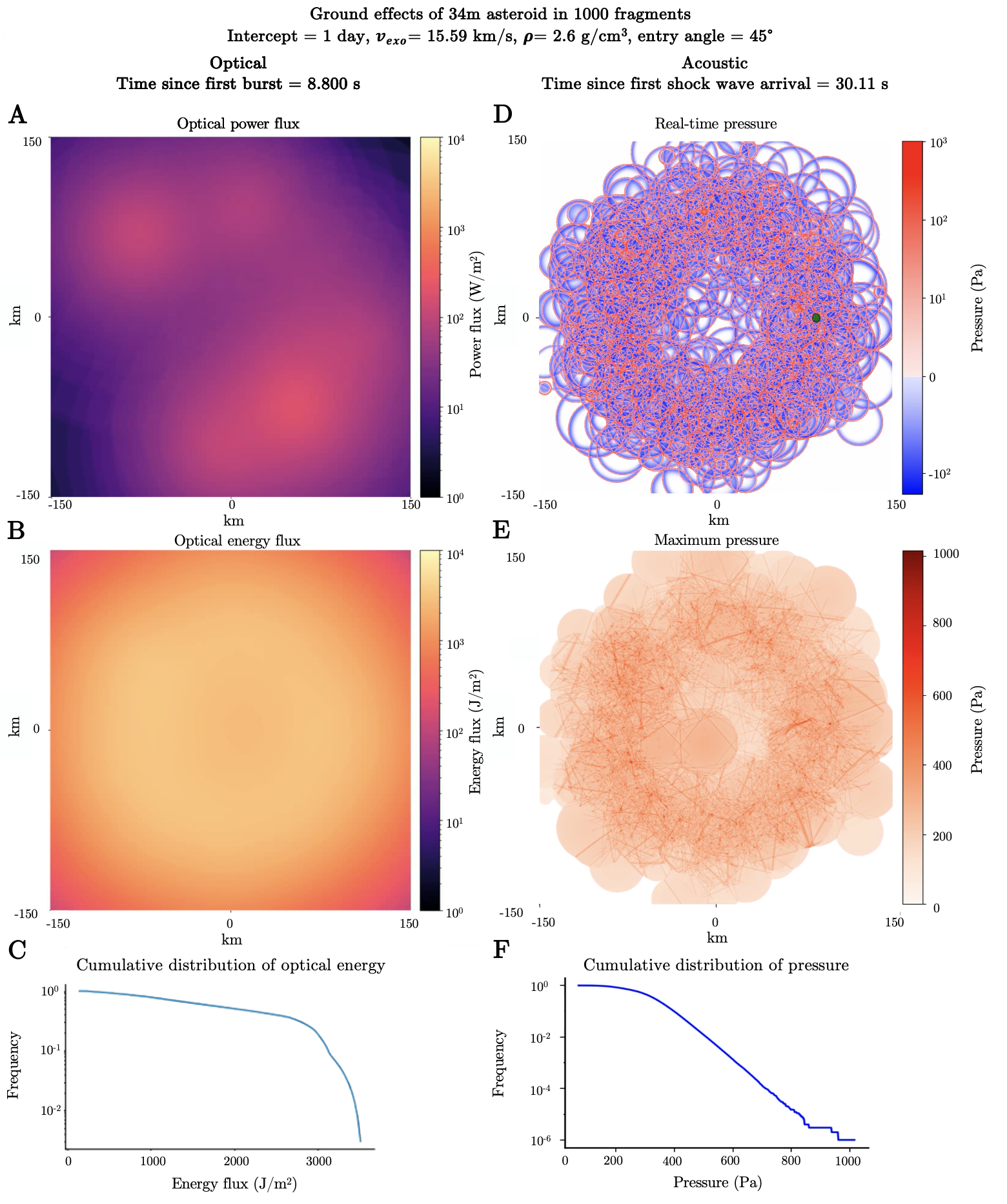}
    \caption{Optical (left) and acoustic (right) ground effects simulations showing a mitigation scenario of a 34 m asteroid (2023 NT1 probable diameter estimate \citep{noauthor_sentry_nodate}) broken into 1000 fragments with a one-day intercept prior to impact. Simulations assume a spherical parent asteroid traveling at 15.59 km/s relative to Earth’s reference frame with an average density of 2.6 g/cm$^3$, entry angle of 45\textdegree, and average fragment disruption of 1 m/s. Note that the current time in each simulation differs. As the optical propagation is occurring at very close to the speed of light in vacuum, and the relevant distance scale from the fragment to the observer are of order of tens to hundreds of km, the real-time of the optical simulation is seen in the title as ``Time since first burst” (units of seconds), dictating the amount of time that has passed since the first airburst of the first fragment. The acoustic simulation instead displays the time since the first shock wave arrived at an arbitrary observer on the ground within the simulation area (as shown by the green dot in Figure \ref{fig:34m GE 1000 in 1 day}D. (A) Real-time optical power flux. (B) Real-time optical energy flux. (C) CDF dictating the frequency of occurrence of various energy flux values. Note that the total optical energy deposition is $\sim$60 times lower than the damage threshold of 200 kJ/m$^2$, and less than 1\% of locations on the ground experience \textgreater 3.5 kJ/m$^2$. (D) Real-time pressure. (E) Maximum pressure experienced in each location throughout the current length of the simulation; each pixel displays the highest pressure it has experienced. Dark orange planes show the caustics (the positive interference from interacting shock waves). (F) CDF dictating the frequency of occurrence of various pressure values. Note that the sum of all shock wave over-pressures (including caustics) is $\sim$2 times lower than the damage threshold of 2 kPa, and less than 1\% of locations on the ground experience \textgreater 0.5 kPa. Higher pressures occur rarely as a result of two-point and three-point caustics. }
    \label{fig:34m GE 1000 in 1 day}
\end{figure*}

\subsection{34 m diameter scenarios: probable diameter estimate} \label{sec:34m}
We highlight several 34 m diameter threat cases as proxies for the probable size estimate for 2023 NT1 \citep{noauthor_sentry_nodate}. For weak rubble-pile and granitic asteroids (average density of 1.4-2.6 g/cm$^3$), we find that a one-day intercept prior to impact with a conservative number of fragments (1000) is sufficient to keep all ground effects below their damage thresholds. Such a mitigation scenario with an average density $\rho$ = 2.6 g/cm$^3$ and entry angle of 45\textdegree\ yields 1\% CDF values of 3.5 kJ/m$^2$ for optical energy flux and 0.5 kPa for shock wave over-pressure (Figures \ref{fig:34m GE 1000 in 1 day} and \ref{fig:34m cdf}; Table \ref{table: frag}). Extremely short time interdiction scenarios also appear feasible; intercept times of 12 hours prior to impact yield reasonable effects (with shorter times likely viable as well) with 1\% CDF values of about 9.0 kJ/m$^2$ and 0.8 kPa (Figure \ref{fig:34m cdf}; Table \ref{table: frag}).

For strong stony to metallic asteroids (average density of 4.0-6.0 g/cm$^3$), one-day intercepts remain feasible (Figure \ref{fig:34m cdf}; Table \ref{table: frag}). A 34 m threat with an average density of 6 g/cm$^3$ and entry angle of 45\textdegree\ disassembled into 1000 fragments with a one-day intercept results in 1\% CDF values of 21.2 kJ/m$^2$ and 1.4 kPa. However, longer intercepts and/or greater fragment numbers are preferred to decrease the magnitude of the ground effects.

For comparison, we estimate that an unfragmented spherical 34 m asteroid with an average density of 2.6 g/cm$^3$ and entry angle of 45\textdegree\ would yield an average optical energy deposition of 80.2 kJ/m$^2$ and average acoustic over-pressure of 1.3 kPa as experienced on the ground, with 1\% CDF values of 357 kJ/m$^2$ and 5.0 kPa, respectively (Figure \ref{fig:34m cdf}; Table \ref{table: unfrag}). If the average density is increased to 6 g/cm$^3$, we find the average ground effects to be 992 kJ/m$^2$ and 3.2 kPa, with 1\% CDF values of about 3800 kJ/m$^2$ and 18 kPa, respectively (Table \ref{table: unfrag}).

\subsection{58 m diameter scenarios: upper limit of diameter estimate} \label{sec:58m} 
\begin{figure*}[!ht]
    \centering
    \includegraphics[width=0.9\textwidth]{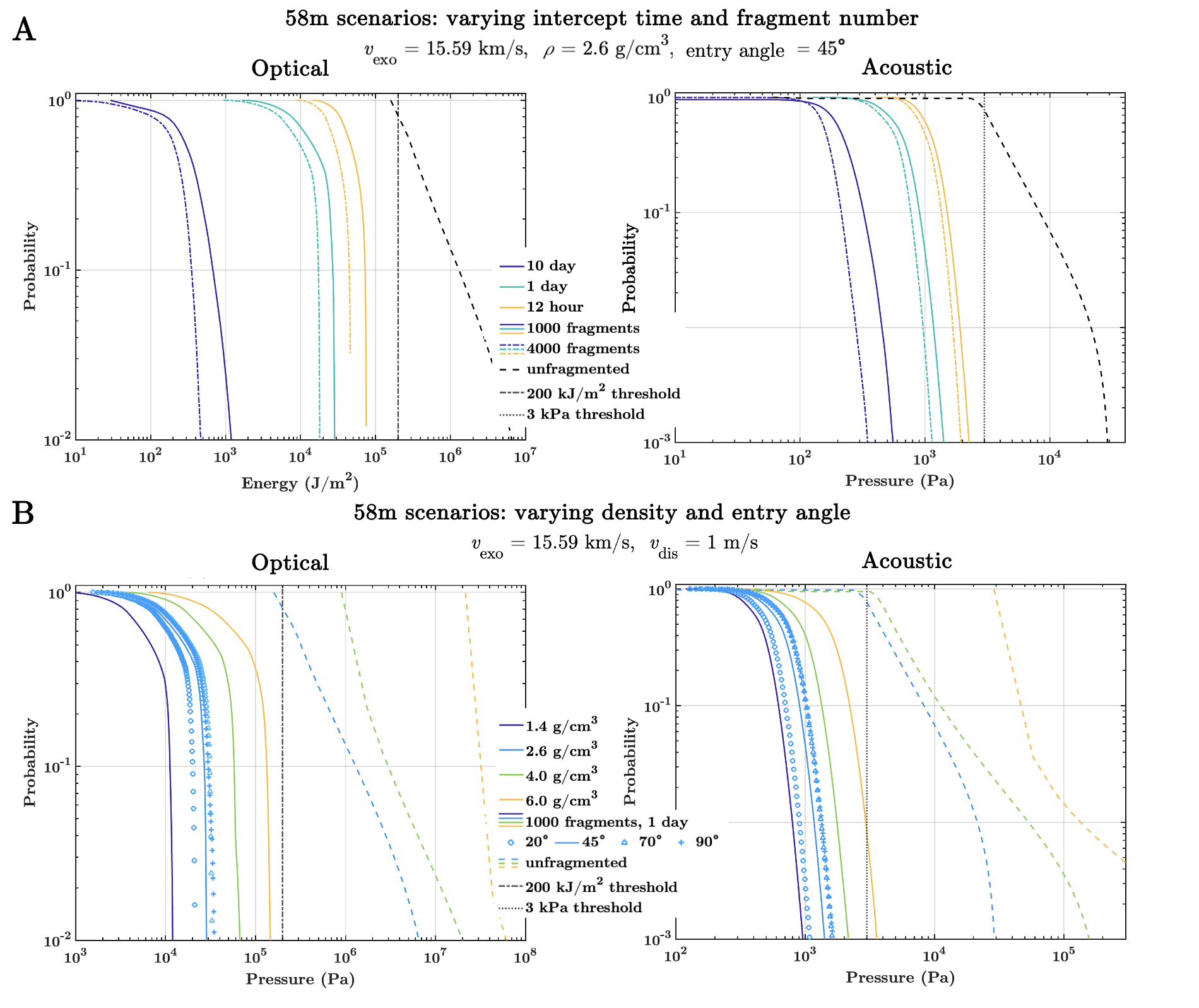}
    \caption{CDFs of the optical (left) and acoustic (right) ground effects of a variety of mitigation scenarios with a 58 m spherical parent asteroid. All scenarios assume an exo-atmospheric parent asteroid velocity (v$_{exo}$) of 15.59 km/s; mitigated scenarios assume an average fragment disruption velocity (v$_{dis}$) of 1 m/s. The dash-dotted black line (left) and dotted black line (right) mark the optical and acoustic damage thresholds of 200 kJ/m$^2$ and 2 kPa, respectively. (A) Varying intercept time and fragment number. All scenarios assume an average density ($\rho$) of 2.6 g/cm$^3$ and entry angle of 45\textdegree\ relative to Earth’s horizon. Legend dictates intercept time prior to impact, number of fragments, and scenario type (fragmented versus unfragmented). (B) Varying density and entry angle. All mitigated scenarios disassemble the parent asteroid into 1000 fragments with a one-day intercept prior to impact. All unmitigated scenarios (dashed lines) assume an entry angle of 45\textdegree\ relative to Earth’s horizon. Unbroken exo-atmospheric energies range from 4.15–17.8 Mt. Legend dictates average density, entry angle relative to Earth's horizon, and scenario type.}
    \label{fig:58m cdf}
\end{figure*}

To simulate higher-risk scenarios of 2023 NT1, we highlight several threats at the high-end diameter estimate of 58 m \citep{noauthor_sentry_nodate}. For weak rubble-pile to granitic asteroids (average density of 1.4-2.6 g/cm$^3$), we again find that disruption into 1000 fragments with a one-day intercept prior to impact is sufficient to keep all ground effects below their damage thresholds. Such a mitigation scenario for a 58 m target with an average density of 2.6 g/cm$^3$ and entry angle of 45\textdegree\ yields 1\% CDF values of 28.7 kJ/m$^2$ for optical energy flux and 1.2 kPa for shock wave over-pressure (Figure \ref{fig:58m cdf}; Table \ref{table: frag}). 

For strong stony to metallic asteroids (average density of 4.0-6.0 g/cm$^3$), one-day intercepts are feasible, although 58 m asteroids with densities above 4.0 g/cm$^3$ may yield ground effects close to, or slightly above, the damage thresholds (Figure \ref{fig:58m cdf}; Table \ref{table: frag}). A 58 m threat with an average density of 6 g/cm$^3$ and entry angle of 45\textdegree\ disassembled into 4000 fragments with a one-day intercept results in 1\% CDF values of 105 kJ/m$^2$ and 2.6 kPa; for a two-day intercept, these effects decrease, with 1\% values of 35.0 kJ/m$^2$ and 1.6 kPa (Table \ref{table: frag}).

For comparison, we estimate that an unfragmented spherical 58 m asteroid with an average density of 2.6 g/cm$^3$ and entry angle of 45\textdegree\ would yield an average optical energy deposition of 1500 kJ/m$^2$ and average acoustic over-pressure of 3.8 kPa as experienced on the ground, with 1\% CDF values of about 6500 kJ/m$^2$ and 21 kPa, respectively (Figure \ref{fig:58m cdf}; Table \ref{table: unfrag}). 

\section{Discussion and future work} \label{sec:discussion}
\subsection{Summary of results}

\begin{figure*}[!ht]
    \centering
    \includegraphics[width=0.9\textwidth]{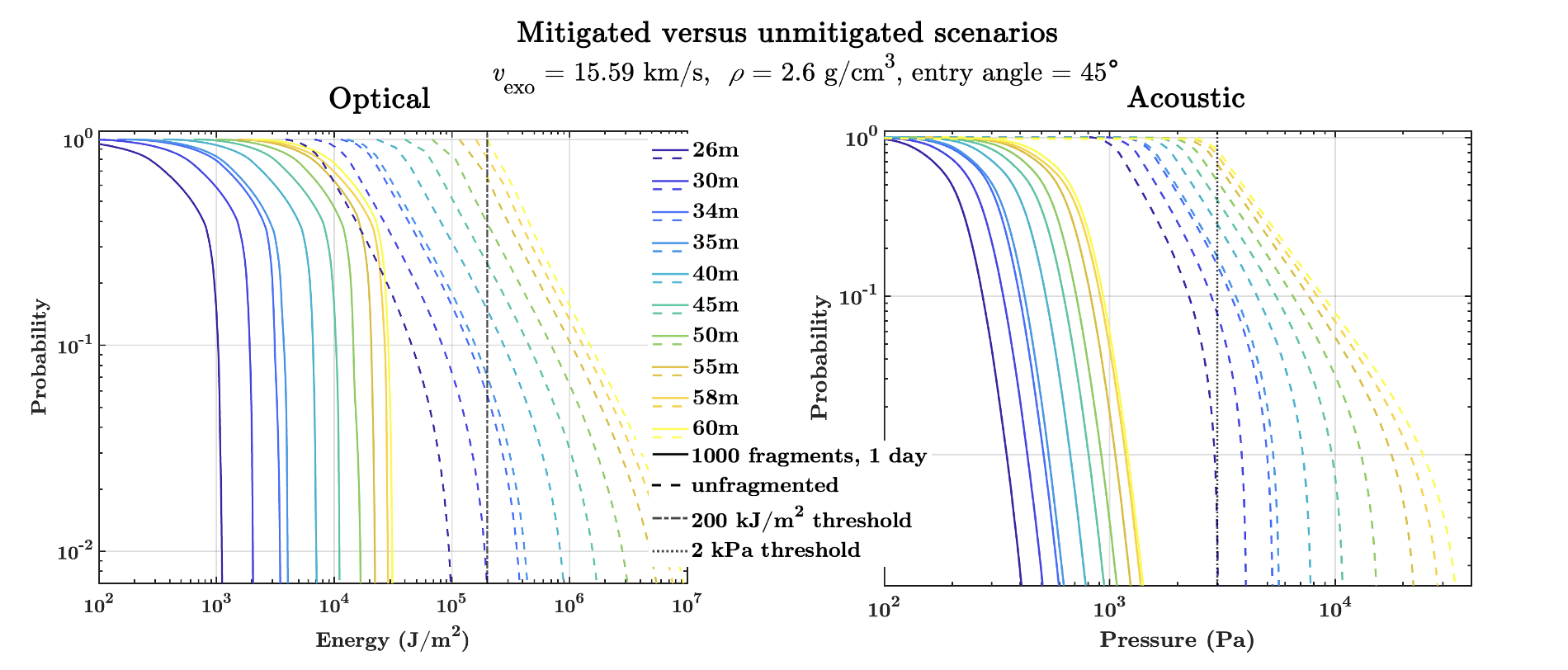}
    \caption{CDFs of the optical (left) and acoustic (right) ground effects for a variety of mitigation scenarios (solid lines) versus unmitigated scenarios (dashed lines). All scenarios assume a spherical parent asteroid with exo-atmospheric velocity (v$_{exo}$) of 15.59 km/s, average density ($\rho$) of 2.6 g/cm3, and entry angle of 45\textdegree\ relative to Earth's horizon; mitigated scenarios disassemble the asteroid into 1000 fragments with a one-day intercept prior to impact and assume an average fragment disruption velocity of 1 m/s. The dash-dotted black line (left) and dotted black line (right) mark the optical and acoustic damage thresholds of 200 kJ/m$^2$ and 3 kPa, respectively.}
    \label{fig:frag vs unfrag}
\end{figure*}

\begin{figure*}[!ht]
     \centering
     \includegraphics[width=0.9\textwidth]{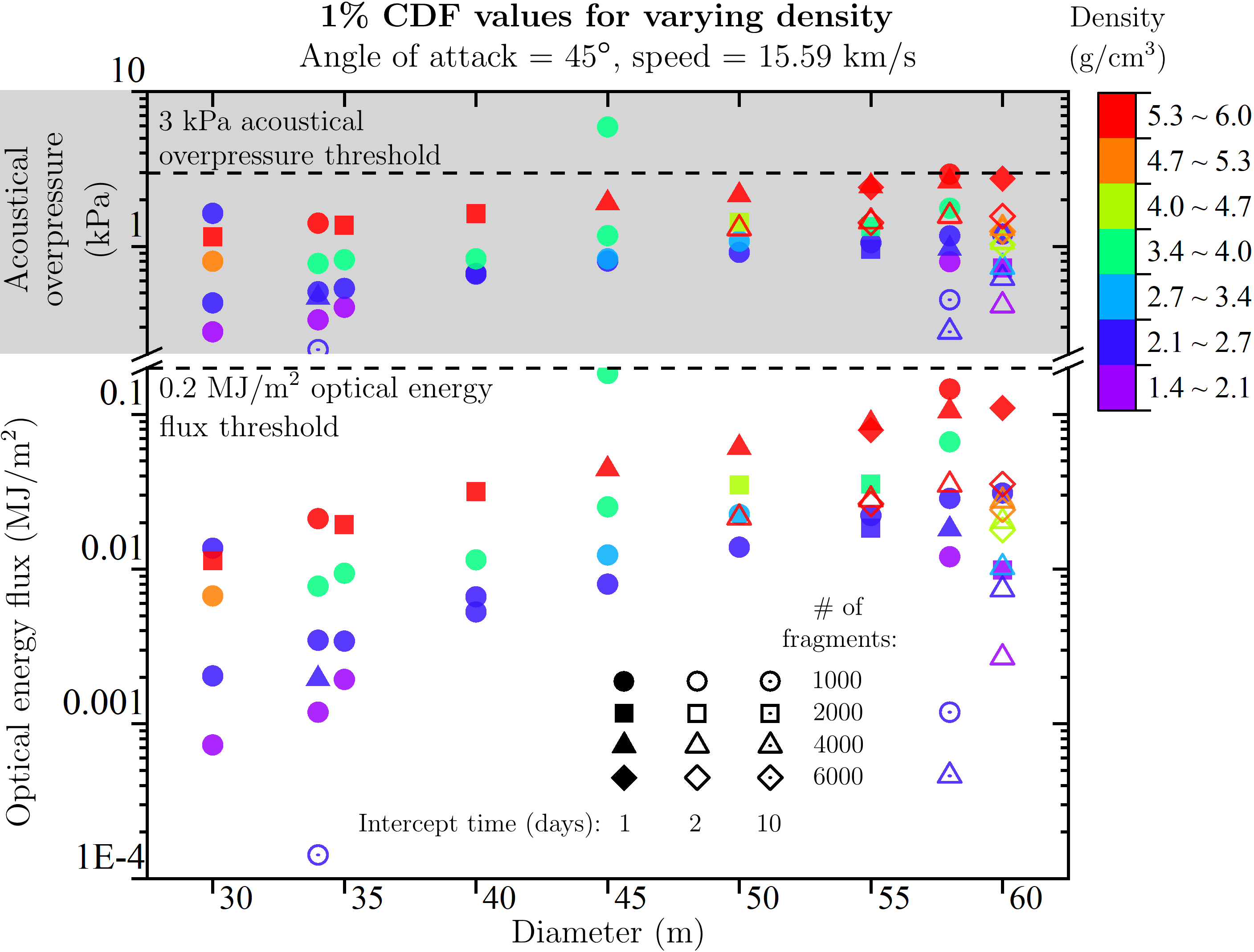}
     \caption{1\% CDF values for optical energy flux and acoustic over-pressure for asteroid 2023 NT1 mitigation scenarios with varying threat diameter ($x$-axis) and density (color scale). Points with filled symbols represent intercepts 1 day prior to impact, open symbols represent 2-day intercepts, and dotted symbols represent 10-day intercepts. The symbol type represents the number of fragments, as shown by the inset legend. Clear trends can be seen which show that reducing the target into a larger number of fragments and/or intercepting it earlier are required to keep the 1\% CDF values below the acceptable damage thresholds as the threat diameter and density is increased.}
     \label{fig:overall cdf}
\end{figure*}

Based on our simulation results, PI appears to be capable of mitigating short-notice threats like asteroid 2023 NT1 for a variety of potential threat characteristics, achieving a reduction in  ground effects when compared to analogous unmitigated threat cases (Tables \ref{table: frag} and \ref{table: unfrag}; Figure \ref{fig:frag vs unfrag}). This is illustrated in Figure \ref{fig:overall cdf}, which shows a comparison of the previously discussed simulation results for both the optical and acoustic ground effects. The 1\% CDF values for optical energy flux (J/m$^2$) and acoustic over-pressure (Pa) are useful metrics for comparing the effectiveness of a given mitigation scenario based on the magnitude of ground effects. The aggregated results, as shown in Figure \ref{fig:overall cdf}, allows for some logical relationships to be resolved and illustrates the importance of considering mission-specific parameters, such as intercept time, impactor mass, and the number of fragments generated, when comparing mitigation scenarios.

\subsection{Limitations of the model}
The results of these ground effects simulations in tandem with the hypervelocity impact simulations suggest that PI may be an effective strategy for mitigating objects like 2023 NT1. However, we acknowledge that our simulation results presented here represent a very limited range of potential threat scenarios, which do not fully encompass a high-fidelity investigation of mitigation via PI for all possible NEOs. Our analogues for 2023 NT1 assume spherical, rubble-pile asteroids within a small size range (20-60 m diameter) and with an impact velocity of 15.59 km/s. While we vary threat parameters such as the average density (1.4-7.0 g/cm$^3$) and entry angle (20\textdegree–90\textdegree), it is important to note that the model presented here spans a limited range of the potential physical and impact parameters for 2023 NT1, or any other possible NEO. While our full suite of simulations (of which only a small fraction are presented within this paper) span asteroid sizes of 15–1000 m in diameter and explore a variety of strength models \citep{lubin_asteroid_2023,bailey_optical_2024}, more work is needed to expand the simulation regime to encompass a wider range of potential threat scenarios.

Additionally, while our simulations suggest that asteroid disassembly via PI could mitigate the threat estimates presented here, this method has not yet been put into practice outside of simulations. Further work is needed to establish physical experiments to probe the testable components of the method (e.g., penetrator design, payload delivery, launch vehicles, hypervelocity impacts on a comprehensive range of threat analogues to verify fragment distributions, etc.) beyond the simulation phase.

\subsection{Limitations of planetary defense}\label{sec:PD limitations}
There is currently no planetary defense system that allows for short-warning mitigation. While it is preferable in any scenario for mitigation to occur as early and as far from Earth as possible, it is reasonable to imagine a case in which early warning is not provided. Thus, a robust planetary defense system would not be limited to a single mitigation strategy but would instead be a highly redundant and layered system, akin to national defense systems. We envision PI as synergistic with existing mitigation strategies, such as deflection, which may be logistically favorable in certain threat scenarios, particularly those with long warning times. 


\subsection{When to deflect, fragment, or accept an impact}
There exists a trade-off between mitigation methods, with different strategies being more favorable depending on the threat scenario.  Deflection methods, such as demonstrated by NASA'S Double Asteroid Redirection Test (DART) \citep{rivkin_double_2021}, can be useful for cases with very long warning times (decades to centuries). For decreased warning times, the disruptive modes of PI could be highly favorable. In comparing deflection to fragmentation, it is important to highlight two key metrics: energy versus momentum transfer and threat scenario. 

Deflection mitigates a threat by transferring momentum, while fragmentation via PI focuses on localized transfer of kinetic energy. These distinct transfer methods yield differences in methodology, execution, and required launch mass. Broadly, the momentum transfer (assuming inelastic collision) applied by kinetic impact is only part of the overall momentum change; studies suggest that the ejecta produced during kinetic impact contributes more momentum than the inelastic collision with the spacecraft alone. For example, DART achieved a momentum enhancement factor of $\beta \approx3.61$ \citep{cheng_momentum_2023}, meaning that the impact transferred roughly 3.61 times greater momentum than if the collision had produced no ejecta at all. However, $\beta$ is highly dependent upon the target asteroid and cannot generally be assumed to be similar in other scenarios \citep{rivkin_double_2021,statler_after_2022}.

\begin{figure}
    \centering
    \includegraphics[width=0.47\textwidth]{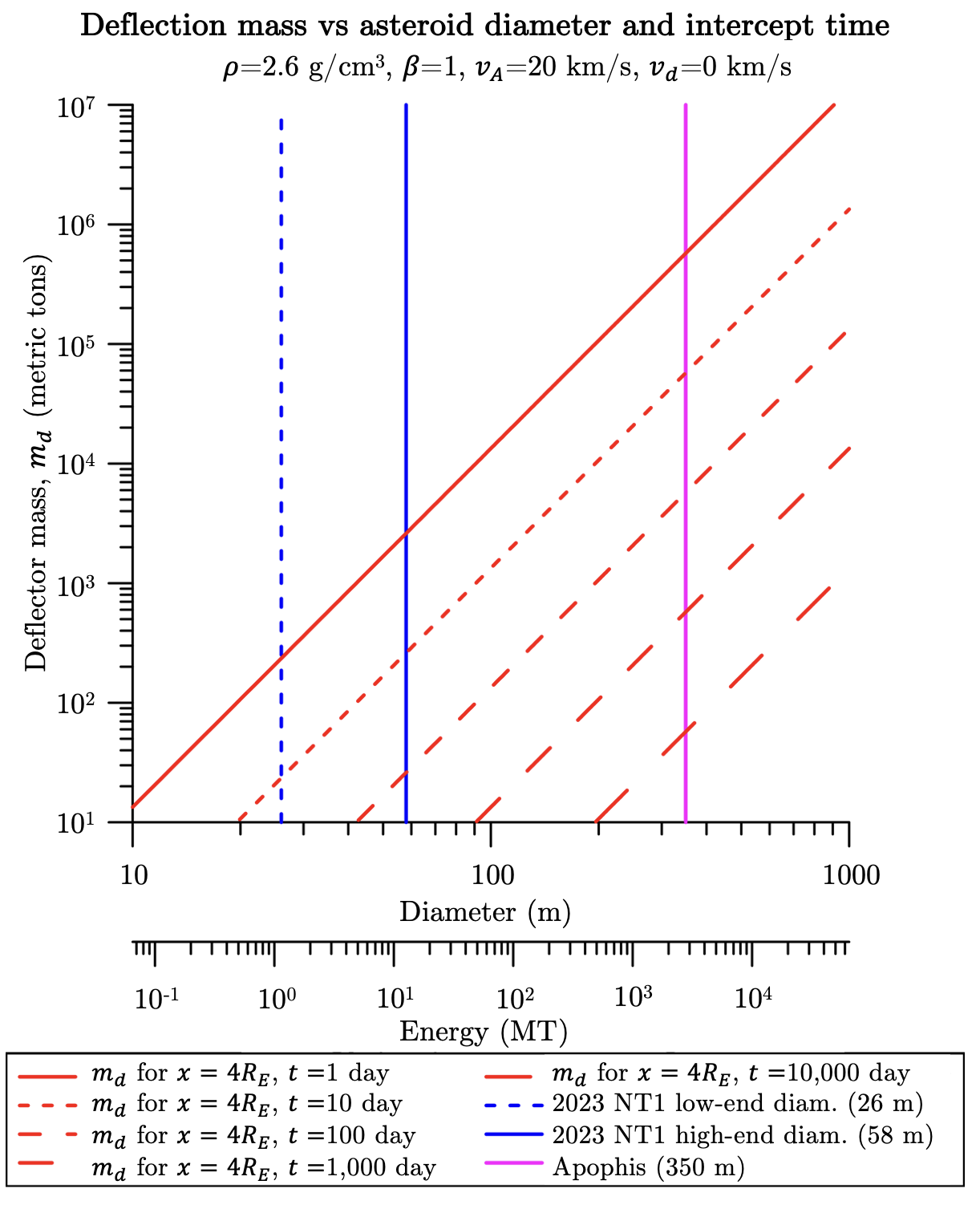}
    \caption{Deflector mass required for a completely inelastic momentum transfer via deflection for threats from 10 m to 1 km diameter for deflection times from 1 day to 10,000 days. Scenarios assume a spherical asteroid with average density ($\rho$) of 2.6 g/cm$^3$ and velocity (v$_{A}$) of 20 km/s and use a deflection miss distance of 4 Earth radii (4R$_{E}$).}
    \label{fig:deflection}
\end{figure}

The choice of the most favorable mitigation method depends heavily on how well the object's physical and orbital parameters are known.  The use of any mitigation method, particularly when used in advance, requires precise observation of the object’s orbital path to determine impact probability and warning time. Focusing our attention on 2023 NT1, objects observed with little to no warning time prior to their closest approach cannot yet be mitigated with certainty. Hypothetically, if 2023 NT1 was observed ahead of time and found to be on a collision course with Earth, how (and whether) to mitigate it would depend upon our knowledge of its characteristics.

For terminal interdiction modes, where warning times can range from mere hours to several days, fragmentation is a more feasible mitigation method. For 2023 NT1’s low-end diameter estimate, we estimate that deflection of a 26 m rubble pile asteroid with an average density of 2.6 g/cm$^3$ and velocity of 20 km/s (relative to Earth) using a one-day intercept prior to impact and a conservative miss distance of 4 Earth radii (x = 4R$_E$) would require approximately 200 metric tons of deflector mass (Figure \ref{fig:deflection}). For the same scenario with a less conservative miss distance of 1 Earth radius (x = R$_E$), we estimate a deflector mass of approximately 50 metric tons. For 2023 NT1’s high-end diameter estimate of 58 m (assuming $\rho=$ 2.6 g/cm$^3$ and v$_A=$ 20 km/s), we estimate the required deflector masses at approximately 3,000 metric tons for x = 4R$_E$ and approximately 750 metric tons for x = R$_E$ for one-day deflection intercepts (Figure \ref{fig:deflection}). For mitigation via fragmentation of the same 26 m and 58 m asteroids with a one-day intercept prior to impact, our simulations suggest the threats could each be effectively disassembled by a single 100 kg penetrator (assuming a closing velocity of 20 km/s) launched aboard any orbital class vehicle of current design.

Note that in order to achieve a scenario with a one-day intercept, the impactor (whether a deflector like DART or a penetrator for fragmentation like PI) would need to be launched several days in advance (dependent upon preferred miss distance). Additionally, the success of deflection is largely dependent upon the deflector’s velocity vector relative to the asteroid’s velocity vector. Assuming the capability to achieve a feasible launch mass for a one-day deflection intercept, a perpendicular impact relative to the direction of travel would likely not provide enough momentum transfer to mitigate the threat.

With warning times on the order of one week or longer for small objects like 2023 NT1, deflection becomes more feasible; however, it may continue to be the case that fragmentation is potentially more effective while requiring less launch mass and potentially shorter launch times prior to impact (Figure \ref{fig:deflection}) \citep{lubin_asteroid_2023}. Using the same 26 m analogue for 2023 NT1 as above, we estimate that deflecting 10 days prior to impact would require deflector masses of approximately 20 metric tons for x = 4R$_E$ and approximately 5 metric tons for x = R$_E$; for the 58 m estimate, we find required masses of approximately 300 metric tons and approximately 75 metric tons, respectively (Figure \ref{fig:deflection}).

In regard to 2023 NT1 and similarly sized threats ($\sim$20–60 m diameter), some may argue that mitigation of asteroids within this size range should be minimal, or perhaps unnecessary, depending on their projected impact location. We introduce a hypothetical scenario in which 2023 NT1 was observed with short warning and estimated to impact a large uninhabited area (e.g., large desert or mid-ocean). It can be argued that such an impact may leave human life and infrastructure unperturbed, thus making a mitigation effort redundant, or perhaps even harmful. It is possible that attempted mitigation of an otherwise potentially harmless impact could yield unintentional consequences; e.g., an unsuccessful deflection effort may push the object towards an inhabited area, or needless fragmentation could spread ground effects over an inhabited area. 

For location-dependent terminal threat scenarios such as these, deciding how and if to mitigate is conditional upon how well the object’s path is known. Given a situation in which there is near certainty that an asteroid will impact an uninhabited area and yield no damage, it may be reasonable to argue that the impact should be allowed to occur unimpeded, or that fragmentation may be risky. However, the justification for such a risk is extremely dependent upon the level of knowledge of the threat. The asteroid’s physical and orbital parameters should be extremely well-constrained, generally from continued and precise observation over time, which may not be possible for terminal scenarios with warning times on the scale of hours to days. The trade-off between impact and terminal mitigation raises questions which have not yet been fully discussed within the field; further work is required to determine the boundary conditions between potential outcomes.

In contrast, we introduce another scenario in which an asteroid is expected to impact an inhabited area that is close to an uninhabited area; e.g., an estimated impact over Los Angeles, California, directly off of the West Coast of the United States. It could be argued that in such a scenario, a small-scale terminal deflection effort may successfully nudge the object into the Pacific Ocean, whereas mitigation via fragmentation could potentially disperse fragments (and thus ground effects) over Los Angeles and/or surrounding inhabited areas. However, our results suggest that the ground effects from terminal-scenario fragmentation of asteroids in the smaller threat regime ($<$200 m diameter) are kept below the estimated damage thresholds mentioned above. Though, we note again that further work is needed to enact physical experimentation of fragmentation, and additional work is needed to determine the limit of justification for mitigation versus unimpeded impact.

\section{Conclusion}\label{sec:conclusion}
The simulations in this study suggest that the Pulverize It (PI) approach for planetary defense is a method worthy of consideration. Capable of operating on both short and extended timelines, PI could offer significant advantages over other mitigation strategies, such as lower launch mass requirements and the ability to mitigate potential threats when given very little forewarning.

For the specific near-miss case of asteroid 2023 NT1, our simulations demonstrate that rubble pile asteroids in the 20-60 meter range could be successfully fragmented, reducing a potential impact threat and the subsequent damages to manageable levels. With relatively minor mass requirements, existing launch vehicles, like the SpaceX Falcon 9, and similar technologies (either preexisting or within reach) could be used to mitigate a wide range of asteroid threats.

Our findings highlight that the PI method could turn a potentially catastrophic impact, like the estimated 1.5 megaton TNT equivalent energy release from asteroid 2023 NT1, into a series of dispersed optical pulses and de-correlated shock waves that would result in minimal effects at ground level. Our simulations estimate that, in scenarios that are designed appropriately for a given threat, it is possible to keep the energy output for the majority of the optical pulses below 200 kJ/m$^2$ and the acoustic over-pressures below 3 kPa. This points to the viability of considering PI as part of a comprehensive planetary defense strategy.

\section{Acknowledgments} 
We gratefully acknowledge funding from NASA Innovative Advanced Concepts (NIAC) Phase I grant 80NSSC22K0764, NIAC Phase II grant 80NSSC23K0966, and NASA California Space Grant NNX10AT93H, as well as from the Emmett and Gladys W. Fund. 
We gratefully acknowledge support from the NASA Ames High End Computing Capability (HECC) group and Lawrence Livermore National Laboratory (LLNL) for the use of their ALE3D simulation tools used for modeling the hypervelocity penetrator impacts and for funding from NVIDIA for an Academic Hardware Grant for a high-end GPU greatly speeding up ground effect simulations. 
Sandia National Laboratories is a multi-mission laboratory managed and operated by National Technology \& Engineering Solutions of Sandia, LLC (NTESS), a wholly owned subsidiary of Honeywell International Inc., for the U.S. Department of Energy’s National Nuclear Security Administration (DOE/NNSA) under contract DE-NA0003525. 
This work was also supported by the Laboratory Directed Research and Development program of Los Alamos National Laboratory (LANL) under project number 20220771ER. LANL is operated by Triad National Security, LLC, for the National Nuclear Security Administration of the U.S. DOE (Contract No. 89233218CNA000001). 
We would like to thank our large team of researchers and undergraduates, including Ruitao Xu, Ysabel Chen, Jasper Webb, Hannah Shabtian, Albert Ho, Alok Thakar, Alex Korn, Kellan Colburn, Daniel Bray, Kavish Kondap, and Jerry Zhang for their tireless work and enthusiasm to defend the planet.


\bibliography{2023NT1_ApJL_bib}{}
\bibliographystyle{aasjournal}

\end{document}